\documentclass[12pt,preprint]{aastex}

\newcommand{\LSI}    {LS I +61 303}

\newcommand{\be}{\begin{equation}}
\newcommand{\ee}{\end{equation}}

\begin{document}

\shorttitle{$\gamma$-ray emission from LS I +61 303
}

\shortauthors{\textsc{Agnieszka Sierpowska-Bartosik \& Diego F. Torres}}

\title{$\gamma$-ray emission from LS I +61 303: \\
The impact of basic system uncertainties}

\author{Agnieszka Sierpowska-Bartosik\altaffilmark{1} \& Diego F. Torres\altaffilmark{1,2}}
\altaffiltext{1}{Institut de Ciencies de l'Espai (IEEC-CSIC)
  Campus UAB, Fac. de Ciencies, Torre C5, parell, 2a planta, 08193
  Barcelona,  Spain. E-mail: agni@ieec.uab.es}
\altaffiltext{2}  {Instituci\'o Catalana de Recerca i Estudis Avan\c{c}ats (ICREA), Spain. E-mail: dtorres@ieec.uab.es }

\begin{abstract}
LS I +61 303 has been recently detected as a periodic $\gamma$-ray source by the Major Atmospheric Imaging Cerenkov (MAGIC) telescope. A distinctive orbital correlation of the  $\gamma$-ray emission was found.
This work shows that the range of uncertainties yet at hand in the orbital elements of the binary system \LSI\  as well as in the possible assumptions on the stellar wind of the optical companion 
play a non-negligible role in the computation of opacities to high energy processes leading to $\gamma$-ray predictions. The geometry influence on the propagation and escape of $\gamma$-ray photons is explored. 
With this study at hand, we analyse the results of a pulsar wind zone model for the production of $\gamma$-rays and compare it with recent MAGIC observations.
\end{abstract}


\section{Introduction}

LS I +61 303 is currently one of the most studied high energy $\gamma$-ray sources (Albert et al. 2006, 2008a,b).  Discovered to shine at TeV energies by MAGIC, it shares with LS 5039 (Aharonian et al. 2005) the quality of being the
only two known mildly variable (Torres et al. 2001) $\gamma$-ray binaries that are spatially coincident with sources above 100 MeV listed in the 
Third Energetic Gamma-Ray Experiment (EGRET) catalogue (Hartman et al. 1999). 
Both of these sources 
show low X-ray emission and variability, and no signs of emission lines or disk accretion (see the recent works by Sidoli et al. 2006,
Chernyakova et al. 2006, Paredes et al. 2007).

Extended, apparently precessing, radio emitting structures at angular extensions of 0.01-0.05 arcsec have been reported by Massi et al. (2001, 2004); but this discovery was not confirmed by recent observations (Dhawan et. al. 2006, Albert et al. 2008a). In fact, Dhawan et al. (2006) presented 
observations from a July 2006 VLBI campaign in which rapid 
changes are seen in the orientation of what seems to be a cometary tail at periastron. This tail is 
consistent with it being the result of a pulsar wind. No large features or high-velocity flows were noted in any of 
the observing days, which implies at least its non-permanent nature. The changes within 3 hours were found to be insignificant, so the velocity 
can not be much over 0.05$c$. 
Albert et al. (2008a) confirmed these finding observing in a different period about a year later, and showed that the 
morphology of the radio emission of the system is maintained. This emphasized the possibility that a pulsar is the compact object companion, since 
the changing morphology of the radio emission along the
orbit would require a highly unstable jet, which details are not expected to be reproduced orbit after
orbit. It is also to be noted that all other Be binaries known to date have neutron-star companions (Negueruela 2004). The reason for this may lie in source evolution if  the rotation of Be stars is achieved during a period of Roche-lobe overflow mass transfer from its initially more massive companion. If the companion loses most of its mass in the process, becoming a He star with mass of only a few M$_\odot$, when exploding as a supernova it can only leave a neutron star behind (see, e.g., Gies 2000, Tauris \& van den Heuvel 2006 and references therein).

There has been a recent burst of activity trying to understand and model the
high energy multi-messenger emission from \LSI, and especially, but not only, assuming that it is a pulsar binary (see, e.g., the latest works by Bosch-Ramon et al. 2006, Bednarek 2006, Gupta and Bottcher 2006, Neronov and Chernyakova 2007, Dubus et al. 2006a,b, Romero et al. 2007, Torres \& Halzen 2007, and Zdziarski et al. 2008). 
The assumptions about the orbital elements and stellar wind properties in these works differ and the impact of their assumptions have generally not been explored. 
%
%
Zdziarski et al. (2008) consider clumpiness of the polar outflow, as a way to explain 
the X-ray variability found on timescales much shorter than the orbital period. Their analysis of the system also finds that the presence of a young pulsar is compatible with all observational constraints. Free-free absorption suppresses most of the radio emission within the orbit, including the pulsed signal of the rotating neutron star. In here, we do not consider clumpiness of the polar wind but rather the impact  (onto  the 
computation of opacities to high energy processes leading to $\gamma$-rays)
of even more basic assumptions, like the uncertainties in the orbital elements of \LSI\ and the main average parameters of the stellar wind of the optical companion. With this study at hand, we analyse the results of a pulsar wind zone model (PWZ) for the production of $\gamma$-rays and compare it with the most recent MAGIC observations.

\section{System uncertainties and assumptions}

\subsection{Orbital elements}

\begin{figure*}[t!]
 \centering
  \includegraphics[width=0.33\textwidth,angle=0,clip]{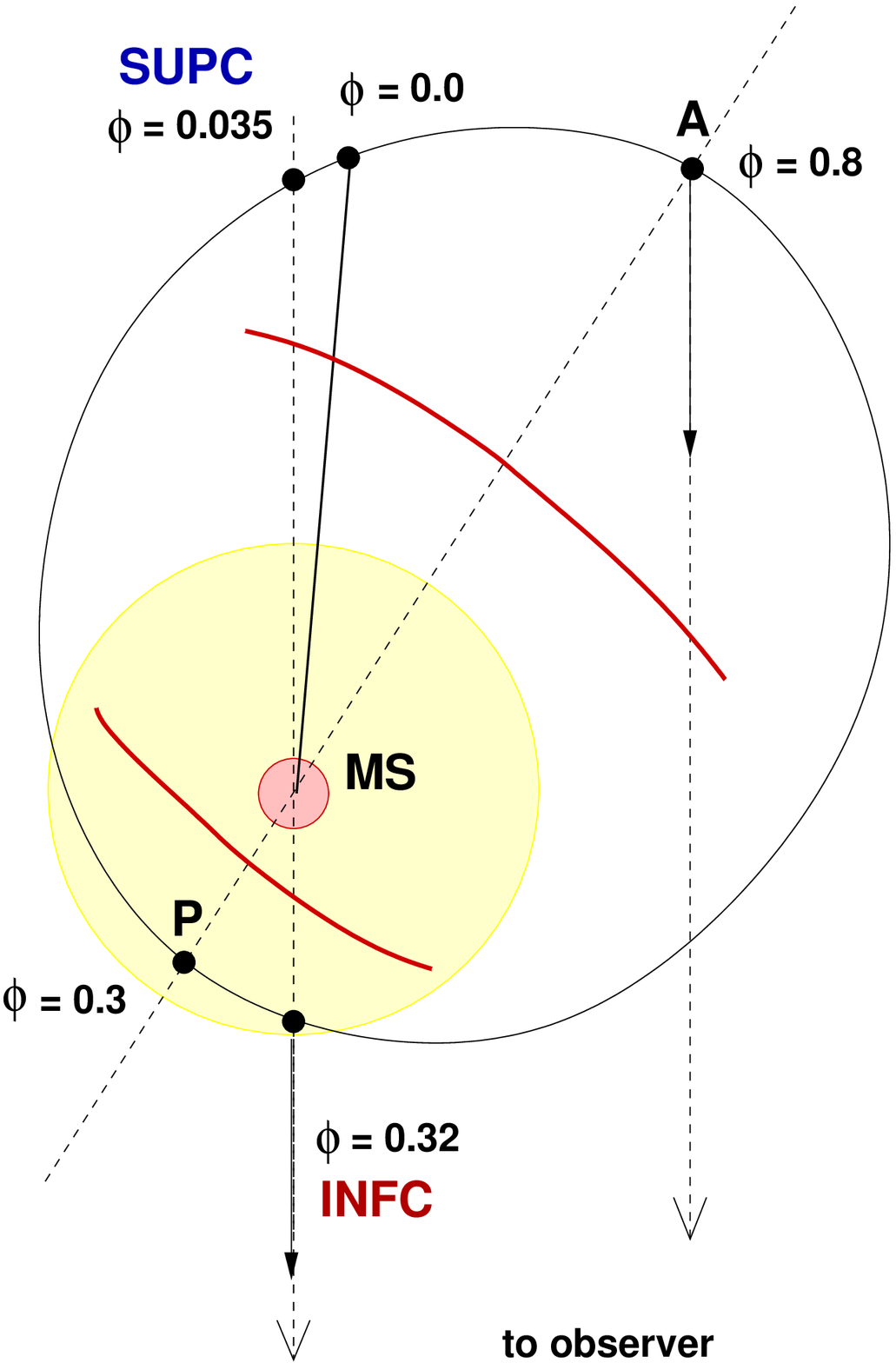}
  \hfill
  \includegraphics[width=0.5\textwidth,angle=0,clip]{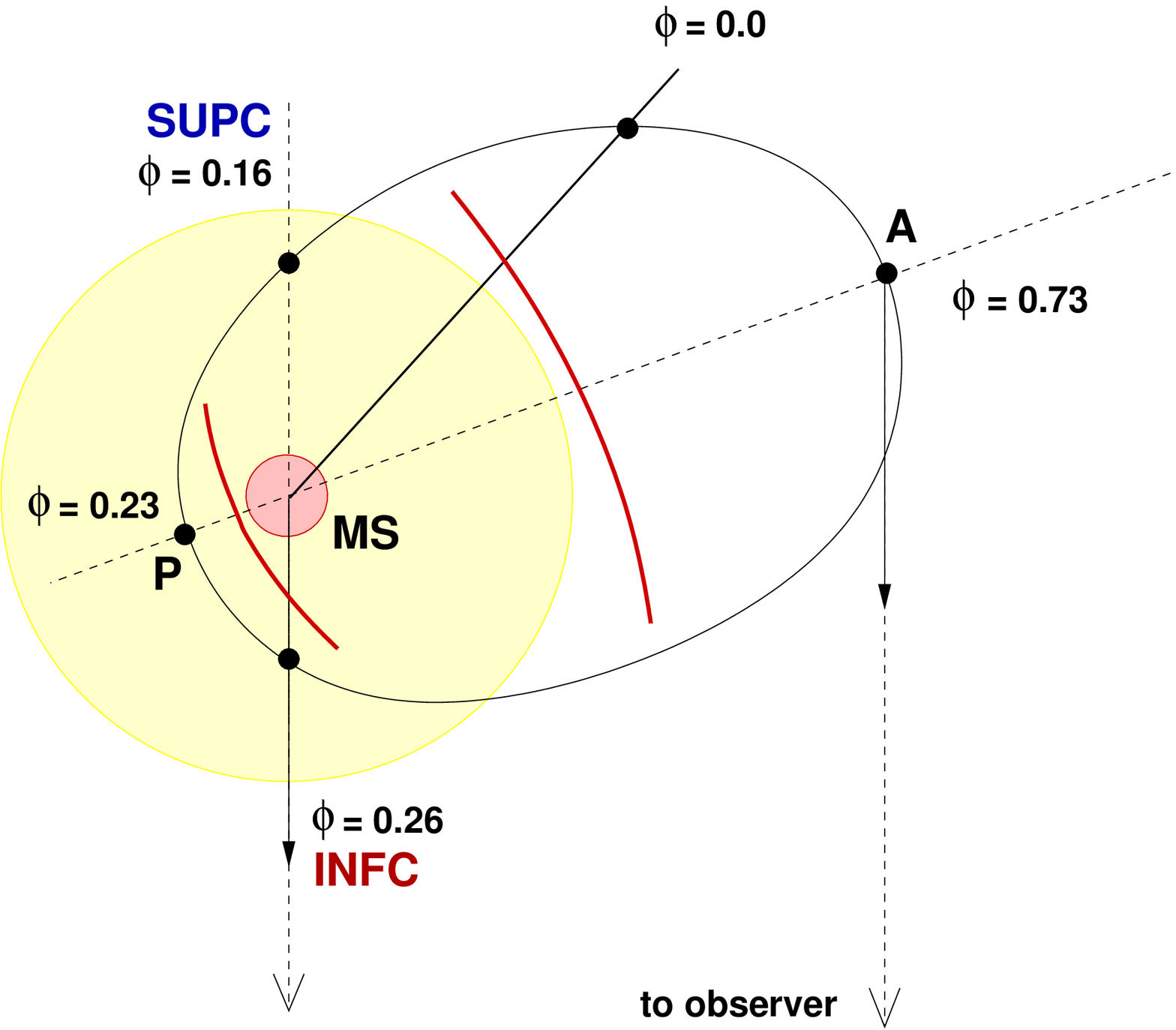}
\caption{\label{fig:geom_obs} {System geometry for two different orbital solutions of LS I +61 303, based on Grundstrom et al. 2007 (left) and Casares et al. 2005 (right). The phases for Inferior conjunction (INFC), Superior conjunction (SUPC), periastron, and apastron are marked accordingly to these solutions. The measurements of phases for this system start from the 
radio ephemeris (see. e.g., Gregory 2002), which exhibits periodic radio outbursts, thus, phase 0 does not correspond to periastron in any of the orbital solutions. The inclination of the system is not shown. Each of the figures are roughly in scale. For the massive star, the equatorial wind (disc) is shown assuming a disc radius $r_d = 7 R_s$. }}
\end{figure*}

The assumed values for the basic binary parameters of \LSI\ needed for our study are summarised in Table \ref{tab:orb-param}. They give account of the two recent solutions for the orbital elements that are available in the literature (Casares et al. 2005, and Grundstrom et al. 2007); { which are schematically shown in Fig. \ref{fig:geom_obs}.}
Whereas they are consistent with each other concerning some of the parameters or   the orbital period; others significantly differ. The nominal values for the eccentricity, when comparing Casares et al. (2005) and Grundstrom et al. (2007), also span a large range, although both values are barely consistent when looking at the given errors: they are quoted as $e=0.55 \pm 0.05$ for the latter and $e=0.72 \pm 0.15$ for the former.
The orbital solution by Grundstrom et al. (2007) seems to be in reasonable agreement with the expectations from the radio model of Gregory (2002),  which places the phase of periastron at $\phi_p=$0.33 -- 0.40. The longitude of periastron { (the angle within the plane of the orbit which defines the position of periastron with respect to the observer; e.g., for $w_{per} = 90^o$, then the periastron is between the star and observer).} significantly
 influences  the geometry of the system and results in different orientation of the binary with respect to the observer. The two quoted values are $\omega_p = 57^o \pm 9$ and $\omega_p = 21.0^o \pm 12.7$ (Grundstrom et al. 2007 and Casares et al. 2005, respectively). Also the difference for the phase of periastron,  $\phi_p=0.301 \pm 0.011$ and $\phi_p=0.23 \pm 0.02$, plays a role in the interpretation of observational data.

Grundstrom et al. (2007) obtained 100 spectra of \LSI, all of them showing the H$\alpha$ emission line and the He {\sc i} $\lambda$6678 feature. To obtain the orbital solution, they assume that the radial velocity variations of this composite profile represent the motion of the Be star, 
since the line formation probably occurs very close to the photosphere of the Be star itself.  Unfortunately, even when the number of observational radial velocity data points have been much increased by Grundstrom et al. (2007), the paucity of measurements in the   orbital phase range 0.4 -- 0.5, and the possibility that the Balmer lines could be contaminated by the disc of the star, still cast some doubts as to what is the real orbital solution of the system. 

\begin{figure*}
  \centering
   \includegraphics[width=0.48\textwidth,angle=0,clip]{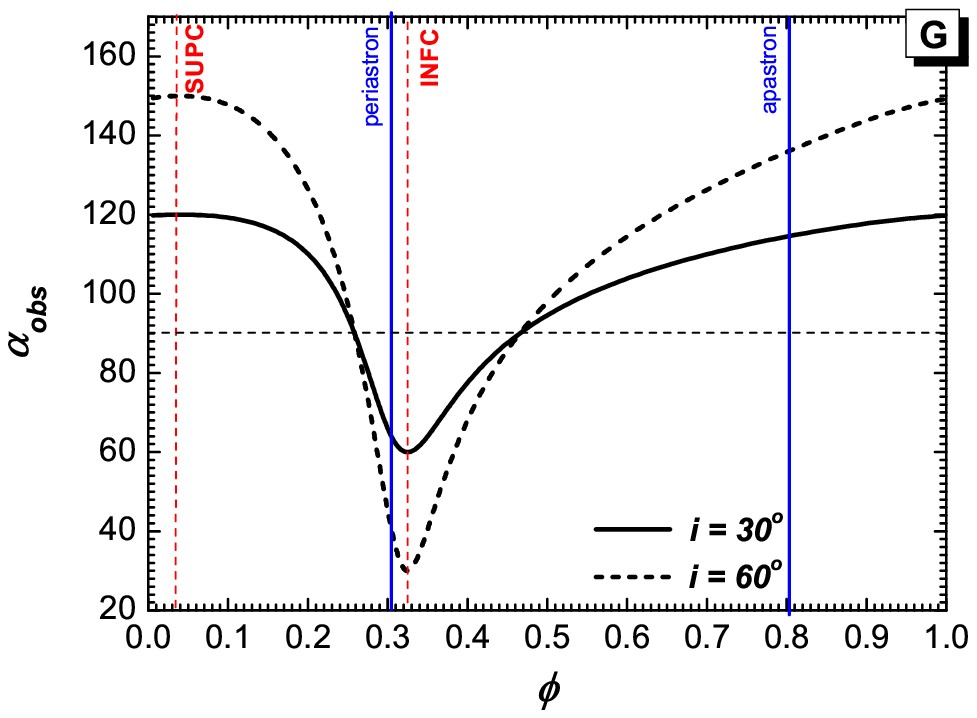}
   \includegraphics[width=0.48\textwidth,angle=0,clip]{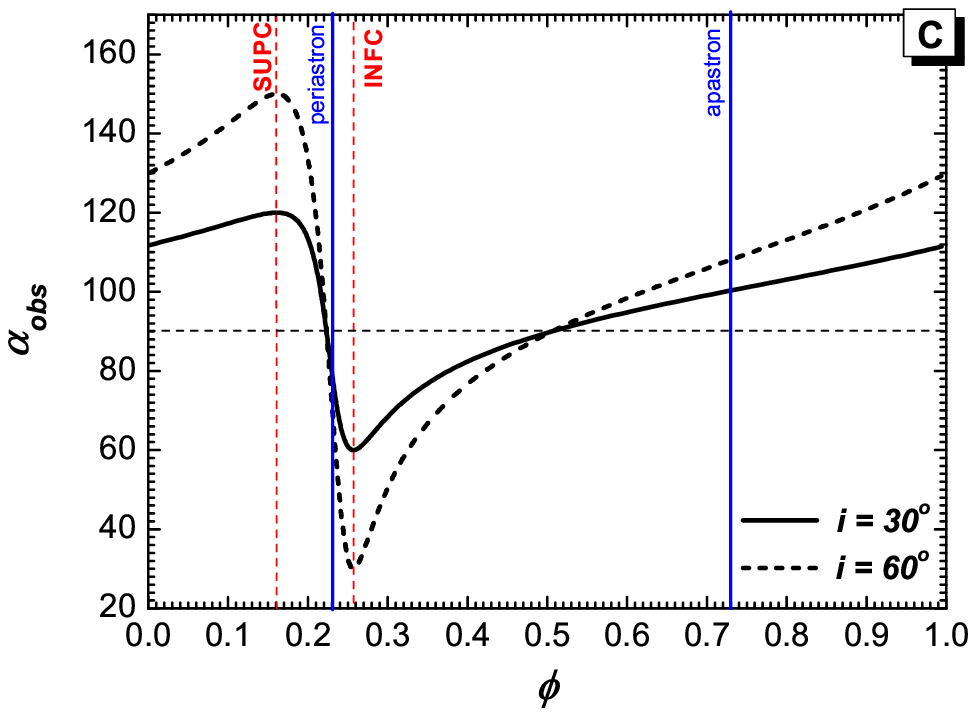}
\caption{The angle to the observer, measured form the pulsar site, as a function of phase along the orbit for two different values of the binary inclination angle $\textit{i}$. The angle was calculated for two different orbital projection, based on Grundstrom et al. 2007 (G - left) and Casares et al. 2005 (C - right). The phases for inferior conjunction (INFC), superior conjunction (SUPC), periastron, and apastron are marked accordingly to the model. The angle is defined to be $\alpha_{obs} = 0$ for propagation outside with respect to the massive star. }
\label{fig:alpha_obs}
\end{figure*}

In what follows, we will work with both sets of orbital solutions and show that such differences, and others given in Table \ref{tab:orb-param}, produce non-negligible changes in the opacities of leptonic processes that lead to high energy emission.
Important geometrical magnitudes, such as the angle to the observer as measured from the pulsar site as a function of phase along the orbit\footnote{See Sierpowska-Bartosik \& Torres (2008) for details on how this angle is computed, and other technical discussions.} are notably affected by the choice of orbital solutions, since they depend directly on the geometry, as can be seen in Fig. \ref{fig:alpha_obs}. In turn, this and other differences unavoidably lead to changes in opacities.

The orbital period of \LSI\ is $\sim 26.5$ days (Taylor \& Gregory, 1982),  with the best measurement coming from the radio observations $(26.4960 \pm 0.0028)$ days (Gregory 2002). 
The massive star is of B0 Ve  type (Hutchings \& Crampton 1981, Paredes et al. 1994). Based on the spectral type (e.g., Cox 2000), its radius and temperature are  $R_s \approx 10 R_{\odot}$ and $T_s = 2.8 \times 10^4$ K, as used by Casares et al. (2005).  If, as in Grundstrom et al. (2007),  we rather assume a radius of $R_s \approx 6.7 R_{\odot}$ (e.g. Harmanec 1988), the effective temperature is $T_s = 3.0 \times 10^4$ K.  A different temperature changes the photon field against which we have computed inverse Compton processes. The  dimensions of the star have also an impact on the presented results for each orbital configuration yielding to the change of relative separation, what is shown in Table \ref{tab:orb-param}; the massive star is not a point-like source, what we have also taken into account.


The estimation of the inclination of the orbital plane is related to the knowledge on the mass of each of the stars. To give an example; from the mass function, $f(M) =0.0107\, M_{\odot}$ (Casares et al. 2005), and taking the possible inclination as $i = 30^o$ or  60$^o$  with the mass of the massive companion in the middle of their preferred mass range $M_s = 12.5 \, M_{\odot}$, one can get the mass of the compact object within $m \sim 2.5 - 1.5 \, M_{\odot}$ (higher inclination implies a lower mass of the compact object, favouring a pulsar scenario). Based on these estimations, we get the semimajor axis of the binary as $a \approx 0.42$ AU ($a \sim 6 \times 10^{12}$ cm). Casares et al. (2005) conclude that the compact object would then be a neutron star for inclinations 25$^o < i <60^o$ and a black hole for $i <25^o$.
Finally, a pulsar scenario for the binary implies an assumption on the spin down luminosity of the pulsar (thus, a free parameter of the model). Typically, for young pulsars it is found in the range $10^{36}-10^{37}$ erg s$^{-1}$. We set this parameter to $L_{sd} = 5 \times 10^{36}$ erg s$^{-1}$ for the following estimations. 

\begin{table*}
\scriptsize
\caption{LS I +61 303: system parameters and a comparison with LS 5039}
\centering
\begin{tabular}{llll}
\hline
Parameter & \multicolumn{2}{c}{Adopted value}  & { LS 5039 }\\
\hline
Distance to the system  $D$ & \multicolumn{2}{c}{$ 2.3 \,\rm kpc$} &  $ 2.5 \,\rm kpc$\\
Semimajor axis  $a$ &  \multicolumn{2}{c}{$6.3 \times 10^{12}$ cm}  & $0.15\, \rm AU$ $\sim 3.5\, R_s$  \\
Mass of star   $M_s$ &  \multicolumn{2}{c}{$12.5\, M_\odot$} & $23\, M_\odot$\\
\hline
     & {Grundstrom et al. 2007} & {Casares et al. 2005} & \\
\hline
Eccentricity of the orbit  $\varepsilon$ & $0.55$ & $0.72$ & $0.35$ \\
Longitude of periastron  $\omega_{p}$ & $57^o$ & $21^o$ & $226^o$ \\
Phase of periastron  $\phi_{p}$ & $0.301$ & $0.23$ & $0$ \\
Periastron separation  $d_p$ & $\sim 6\, R_s$ & $\sim 2.6\, R_s$ & $\sim 2.3\, R_s$ \\
Apastron separation  $d_a$ & $\sim 21\, R_s$ & $\sim 15.7\, R_s$ & $\sim 4.7\, R_s$ \\
Radius of star   $R_s$ & $6.7\, R_\odot$  & $10.0\, R_\odot$  & $9.3\, R_\odot$\\
Temperature of star  $T_s$ & $ 3.0 \times 10^4$ K& $ 2.8 \times 10^4$ K& $3.9 \times 10^4$ K\\
\hline
Mass loss rate of star (polar wind)  $\dot{M}_p$ &  \multicolumn{2}{c}{$10^{-8}\, M_\odot\, \rm yr^{-1}$}  & $10^{-7}\, M_\odot\, \rm yr^{-1}$  \\
Wind termination velocity (polar wind)  $V_{\infty}^{pol}$ &  \multicolumn{2}{c}{$2000 \,\rm km\, s^{-1}$} & $2400\,\rm km\, s^{-1}$ \\
Wind initial velocity (polar wind)  $V_0^{pol}$ &  \multicolumn{2}{c}{$20\,\rm km\, s^{-1}$} &  $4 \,\rm km\, s^{-1}$ \\
Mass loss rate of star (equatorial wind 1)  $\dot{M}_d$ &  \multicolumn{2}{c}{$1.3 \times 10^{-7}\, M_\odot\, \rm yr^{-1} $}  & -  \\
Mass loss rate of star (equatorial wind 2)  $\dot{M}_d$ &  \multicolumn{2}{c}{$1.3 \times 10^{-6}\, M_\odot\, \rm yr^{-1} $}  & -  \\
Wind termination velocity (equatorial wind)  $V_{\infty}^{d}$ & \multicolumn{2}{c}{$300 \,\rm km\, s^{-1}$} & - \\
Wind initial velocity (equatorial wind)  $V_0^{d}$ & \multicolumn{2}{c}{$5\,\rm km\, s^{-1}$} & - \\
Rotational velocity of the star  $V_{rot}$ & \multicolumn{2}{c}{$400 \,\rm km\, s^{-1}$} & - \\
The half opening angle of the disc  $\theta_d$ & \multicolumn{2}{c}{$15^o$} & - \\
The radius of the disc  $r_d$ & \multicolumn{2}{c}{$7$ R$_s$} & - \\
\hline
\vspace{0.2cm}
\end{tabular}
\label{tab:orb-param}
\end{table*} 

\subsection{Stellar wind}

\begin{figure*}[t]
  \centering
   \includegraphics*[width=0.48\textwidth,angle=0,clip]{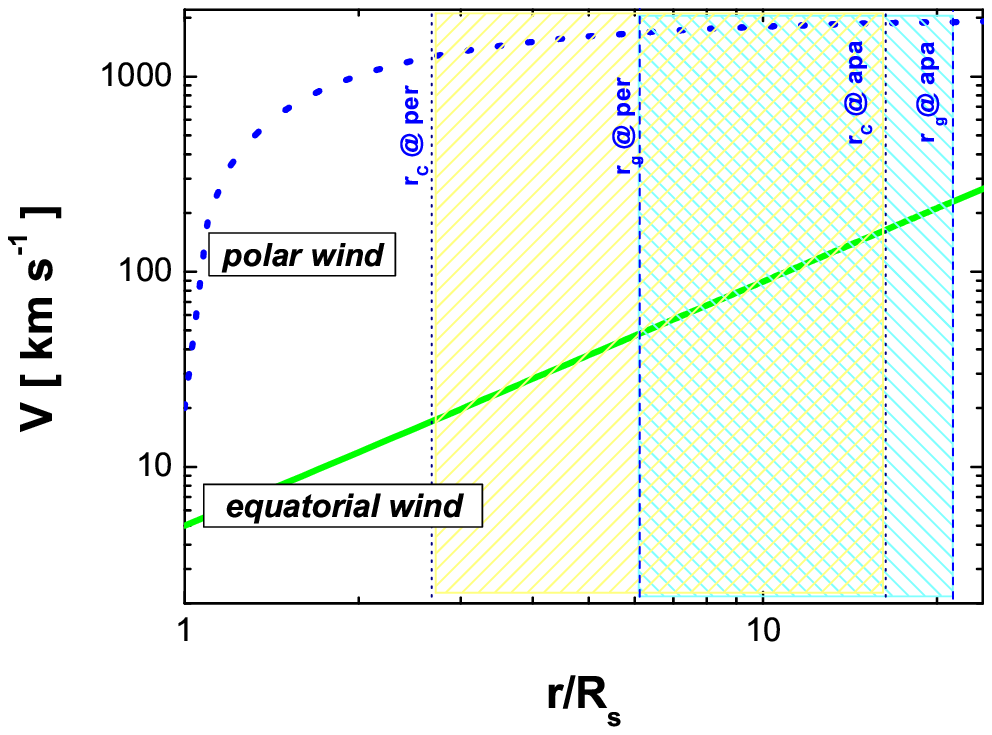}
   \includegraphics*[width=0.48\textwidth,angle=0,clip]{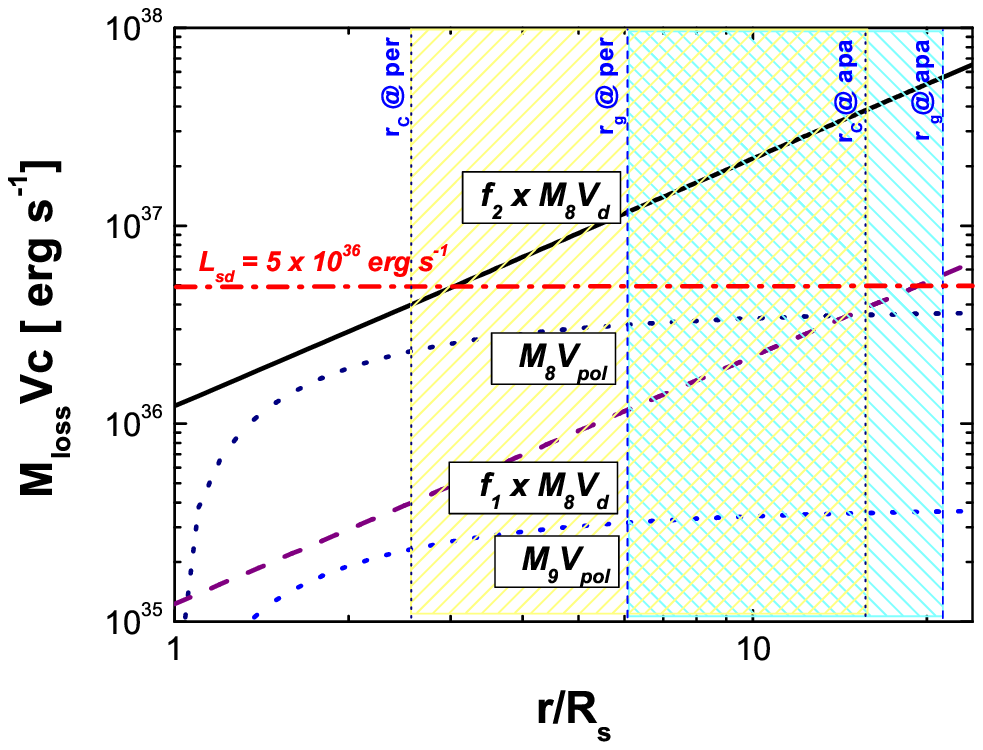}
\caption{\label{fig:Vpd_MVc} {Left: the velocities of the two components of the massive star wind as a function of distance from the star centre. The parameters of the wind are in Table \ref{tab:orb-param}. Right: kinetic power of the polar and equatorial massive star winds versus radius (from the centre of the massive star). For polar wind the power is calculated for mass loss rate $M_8 = \dot{M}_p = 10^{-8} \, M_\odot\, \rm yr^{-1}$ and $M_9 = \dot{M}_p = 10^{-9} \, M_\odot\, \rm yr^{-1}$, for comparison. For the equatorial wind the power is calculated for the scaling factors $f_1 = 50$, $f_2 = 500$ and $M_8$ for both cases. The shaded areas correspond to the range of the binary separation along the orbit (from periastron to apastron) for two models of the binary: based on Grundstrom et al. 2007 (light blue) and Casares et al. 2005 (light yellow). The distances which correspond to the separation at the periastron and apastron for each model are marked with vertical blue lines. The pulsar spin down power is also marked. }}
\end{figure*}

The stellar wind of the Be companion is assumed to have two components, one related with a polar contribution, and the other, with the equatorial disk (e.g. Waters et al. 1988).
The polar wind is radiatively driven (e.g. Castor \& Lamers 1979), for which the velocity law is   
$
 V_p(r) = V_o^{pol} + (V_{\infty}^{pol} - V_o^{pol}) \left(1 - {R_s}/{r}\right)^{\beta},
\label{eq:vp}
$
and where typically, $\beta \approx 1$, $V_o^{pol} = 0.01 V_{\infty}^{pol}$, and the terminal velocity of the wind $V_{\infty}^{pol} = 1500-2000 \,\rm km\, s^{-1}$, with mass loss rates obtained from UV resonance lines: $\dot{M}_p \sim 10^{-8}\, M_\odot\, \rm yr^{-1}$ (e.g., Snow 1981, Waters et al. 1988). 
For the equatorial wind  (e.g., see Waters 1986, Waters et al. 1988, Gregory \& Neish 2002), the velocity law is 
$
V_{eq}(r) = V_o^{eq} \left( {r}/{R_s}\right)^{m}
\label{eq:vd}
$
where $m = 1.25$ (from the density profile assumption for LS I +61 303, see Waters et al. 1988) and $V_0^{eq} = 5\,\rm km\, s^{-1}$.  These authors further assume  that the terminal velocity of the equatorial wind is $V_{\infty}^{eq} \sim \rm few\, 100\, km\, s^{-1}$, so that $V_{\infty}^{eq} \ll V_{\infty}^{pol}$, an assumption mimicked by many other works in the field. The velocity laws of both the polar and equatorial outflows are depicted in the left panel of Fig. \ref{fig:Vpd_MVc}.

The mass loss rate for the equatorial wind in Be stars is generally obtained from IR excess and H$\alpha$ lines. For  \LSI\ it was quoted as  $\dot{M}_d = (1-4) \times 10^{-7}\, M_\odot\, \rm yr^{-1}$ by Waters et al. (1988), although uncertainties in this value exist, since it depends on additional knowledge of the disc half opening angle and initial wind velocity.
A relationship between the mass fluxes of the two wind regions is (e.g., Lamers \&Waters 1987) 
$
 F_d / F_p \approx  \dot{M}_{IR}  / (\dot{M}_{UV} \sin \theta),
$
where $F_d / F_p$ is the ratio of the mass fluxes of the equatorial and polar regions, respectively. Using  $F_d = \dot{M}_{IR}  / (4 \pi r^2 \sin \theta)$ and 
 $F_p = \dot{M}_{UV}  / (4 \pi r^2)$, with an opening angle of 15$^o$, typical values are $\dot{M}_d = (10-10^3) \dot{M}_p \sin \theta$.
In what follows, for the models investigated here we use the description
$
 \dot{M}_d =  f \dot{M}_p  \sin \theta,
$
where $f $ is the specific scaling factor defining the model.
The difference between the mass loss rates of the polar and disc regions should be significant to have a luminosity in X-rays as observed in most of Be stars. But in the case of \LSI\, as there is no accretion signatures, this value is difficult to set. In addition, based on results from X-ray and IR observations,  it was found that the densities of the Be star disc region are higher than in the case of isolated Be stars (e.g., Reig et al. 2000). An accompanying neutron star can trim the disc, preventing further growth, and therefore making it denser. Estimations of the disc radius from H$\alpha$ lines have been made by Grundstrom et al. (2007), finding it in the range of  $\sim 5R_s$. 
In what concerns the high energy $\gamma$-ray emission, the value of the disc truncation is not a trivial parameter.  The smaller the radius at which the equatorial wind is terminated, the more time the pulsar spend outside this internal disc. In the following, we assume the disc radius $r_d \sim 7 R_s$, to account for the fact that the disc can occasionally grow to larger radius (e.g., see Grundstrom et al. 2007, also Zdziarski et al. 2008).  Still, the disc size is much smaller than typical values found for isolated Be stars. The pulsar is then subject to the equatorial wind only during part of the orbital period, and we consider it subject to the polar component otherwise. In Table \ref{tab:phase} we put the characteristic phase and angles (measured from the periastron), where notation ``disc IN'' corresponds to the phase when pulsar enter into the equatorial wind region, and ``disc OUT'' when it leaves this region.
Intrinsic to the uncertainties in these parameters is the fact that for 
such wide choice of polar/disc features we may have a variety of possible situations: the polar wind mass flux can be {\em stronger} than the equatorial one, or viceversa. To complicate things, the ratio of mass fluxes may change with separation as is due to different velocity laws in the equatorial and polar winds.

\begin{table}
\caption{LS I +61 303: characteristic orbital phases in different geometries}
\centering
\begin{tabular}{lllll}
\hline
phase  &  \multicolumn{2}{c}{Grundstrom 2007} &  \multicolumn{2}{c}{Casares et al. 2005} \\
\hline
 & angle $[^o]$ & phase & angle $[^o]$ & phase \\
\hline
SUPC & $213^o$ & $0.035$ & $249^o$ & $0.161$ \\
periastron & $0^o$ & $0.301$ & $0^o$ & $0.23$ \\
INFC & $33^o$ & $0.324$ & $69^o$ & $0.257$ \\
apastron & $180^o$ & $0.801$ & $180^o$ & $0.73$ \\
\hline
disc IN & $311^o$ & $0.265$ & $239^o$ & $0.141$ \\
disc OUT & $49^o$ & $0.337$ & $121^o$ & $0.319$ \\
\hline
\vspace{0.05cm}
\end{tabular} 
\label{tab:phase}
\end{table}

The fact that the pulsar is not in the disk region all the time give rise to mixed wind models, where the influence of polar and disk wind regions must change along the orbit. To explore this effect, we assume that the difference in the mass loss rates is in the range of $f \sim 50 - 500$ what gives (together with the assumption for half opening angle $\theta = 15^o$) the mass loss rates themselves as $\dot{M}_d \sim 13 - 130 \times \dot{M}_p$. These values are summarised in Table \ref{tab:orb-param}, where they are referred to as model 1 ($f=50$) and 2 ($f=500$).
In the right panel of Fig. \ref{fig:Vpd_MVc} we see how the power of the massive star wind changes with respect to the assumed constant pulsar power along the orbit. Depending on $f$, we get two quite different scenarios as the power of the equatorial wind for model 1 ($f = 50$) is lower then the pulsar spin down power only for large distances from the massive star, while for model 2 ($f = 500$) it is larger than $L_{sd}$ already at a few stellar radius. We remark that these scenarios are all within current uncertainties in the knowledge of the system.

\subsection{Hydrodynamic balance}

\begin{figure*}
  \centering
   \includegraphics[width=0.48\textwidth,angle=0,clip]{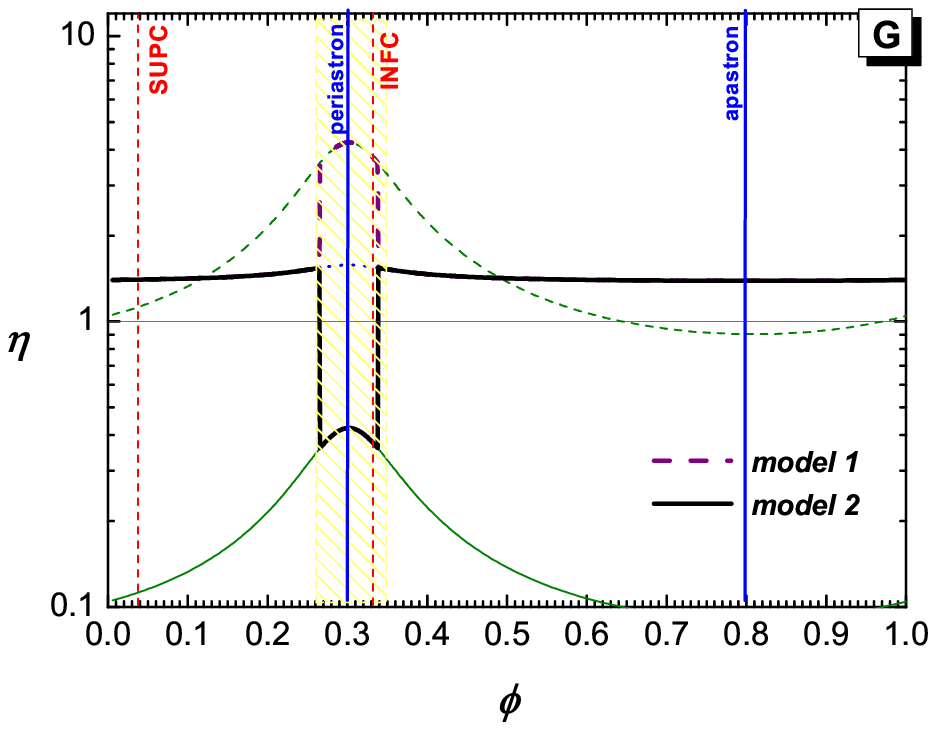}
   \includegraphics[width=0.48\textwidth,angle=0,clip]{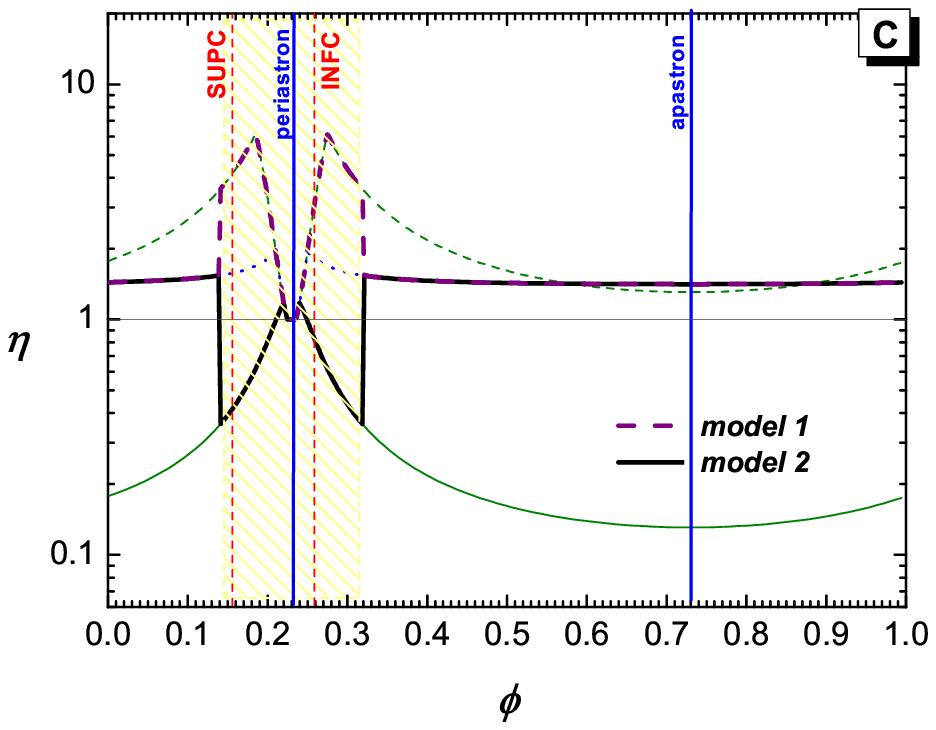}\\
   \includegraphics[width=0.48\textwidth,angle=0,clip]{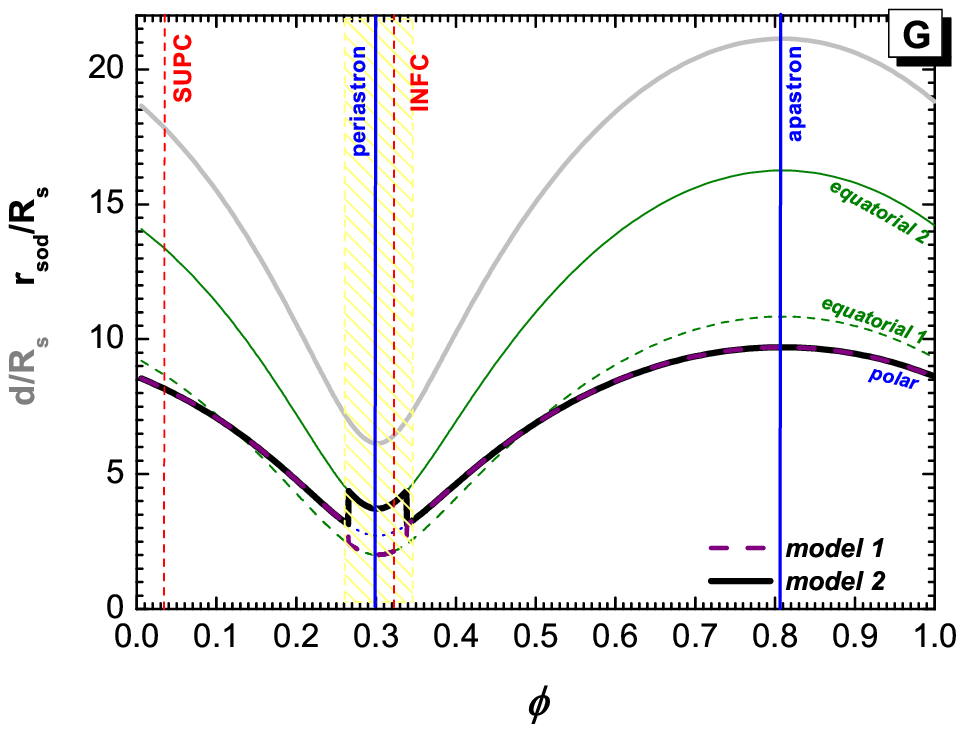}
   \includegraphics[width=0.48\textwidth,angle=0,clip]{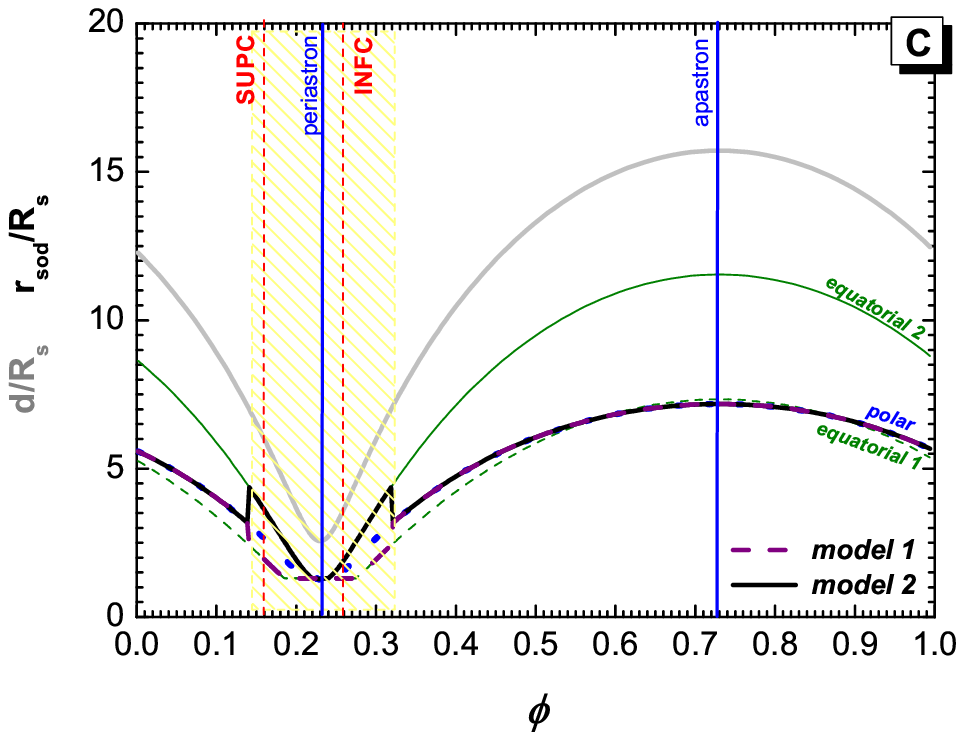}
\caption{\label{fig:etaGC} {Top panel: The $\eta$ parameter for two models of the \LSI\ binary, based on Grundstrom et al. 2007 (G - left) and Casares et al. 2005 (C - right), model 1 ($f = 50$, dashed line), model 2 ($f = 500$, solid line). As a comparison the dependencies for the polar and two equatorial winds (see Table \ref{tab:orb-param}) are shown (blue dotted line for polar wind and thin green dashed and solid lines for equatorial wind in model 1 and 2, respectively ). See the text for a full explanation.
Bottom panel: The stand-off distance of the shock (measured form the massive star centre), $r_{sod}$, in units of stellar radius $R_s$, as a function of binary phase for two models of the system (G, C).}}
\end{figure*}

Based on the hydrodynamic equilibrium of the flows, the geometry of the termination shock is described by the parameter 
$
\eta = {\dot{M}_i V_i}/{\dot{M}_o V_o} ,  
$%
where $\dot{M}_i V_i$ and  $\dot{M}_o V_o$ are the loss mass rates and velocities of the two winds (Girard and Wilson, 1987). If one of the stars is a pulsar of a spin down luminosity $L_{sd}$ and the power of the massive star is $\dot{M_s} V_s$, the parameter $\eta$ can be calculated from the formula
$
\eta = {L_{sd}}/({c(\dot{M_s} V_s)})
$
(e.g., Ball and Kirk 2000).  { This hydrodynamical construct only fixes the shock position, and since we study PWZ processes in this paper, the current approximation is sufficient: In this approach, the shock morphology is not dealt with and plays no significant role in the result.} Note that for $\eta < 1$, the star wind dominates over the pulsar's and the termination shock wraps around it. Note also that for $\eta = 1$, the shock is at equal distance, $d/2$, between the stars. As a first approximation, the shock will be symmetric with respect to the line joining two stars, with a shock front at a distance $r_s  $ from the pulsar
$
r_s  = D  {\sqrt{\eta}}/{(1+\sqrt{\eta})}. 
\label{r0}
$
The surface of the shock front can be approximated then by a cone-like structure with opening angle given by 
$
 \psi = 2.1 \left(  1 - {\overline{\eta}^{2/5}} /  {4} \right) \overline{\eta}^{1/3},
$
where $\overline{\eta} = min(\eta, \eta^{-1})$. This last expression was achieved under the assumption of non-relativistic winds in the simulations of the termination shock structure in Girard and Wilson (1987), albeit it is also in agreement with the relativistic winds case (e.g., Eichler and Usov, 1993; Bogovalov et al., 2007). 

$\eta$-values for the assumed magnitudes of the \LSI\ system are depicted in Fig. \ref{fig:etaGC} (top panels).
The  values calculated in the case of a polar wind only are marked by a thin dotted line (mostly covered by thicker model lines, since the pulsar is out of the disc most of the orbit).\footnote{The description that follows corresponds to the on-line version of the figure.} Assuming the parameters of  the equatorial wind only, the $\eta$-values are marked by thin dashed (green) line for model 1 with $f = 50$, and thin solid (green) line if $f = 500$, these are also plotted beyond their $\phi$-range of validity to facilitate interpreting how the mixed models 1 and 2 are constructed.
With the disc radius $r_d = 7 R_s$ and the half-opening angle $\theta_d = 15^o$ the phases when the pulsar passes trough the disc region are shown as shaded area. 
We can see that the shock front changes position between the stars with the orbital phase (depending on the separation), and that $\eta$ for different models can be larger or smaller than 1. In general (for both binary orbital scenarios), the wind interaction in the polar wind regions gives $\eta > 1$ meaning that the shock is closer to the massive star than to the pulsar. It can be seen that assuming a given power for the equatorial wind (not only the pulsar's) is essential to any model.

Regarding spatial information, we see that for different parameters of the equatorial wind, $\eta$ changes such as for model 2, the PWZ is terminated closer to the pulsar, see the bottom panels in Fig. \ref{fig:etaGC} which show the stand-off distance corresponding to those scenarios (the stand-off distance is independent on the binary inclination as it is measured along the separation axis from the massive star centre). In these bottom panels, we mark results corresponding to 
model 1 ($f = 50$) with a dashed line and to model 2 ($f = 500$) with a solid line.  The stand-off distance calculated in the case of polar wind only (isotropic wind) is marked by thin dotted line (again mostly covered by model lines, since the pulsar spends most of the orbit out of the disc). Assuming only the parameters of  the equatorial wind the stand-off distance is marked by a thin dashed (green) line for the model with $f = 50$, and with a thin solid (green) line if $f = 500$. INFC, SUPC, periastron, and apastron phases are marked. 

\begin{figure*}
  \centering
   \includegraphics[width=0.48\textwidth,angle=0,clip]{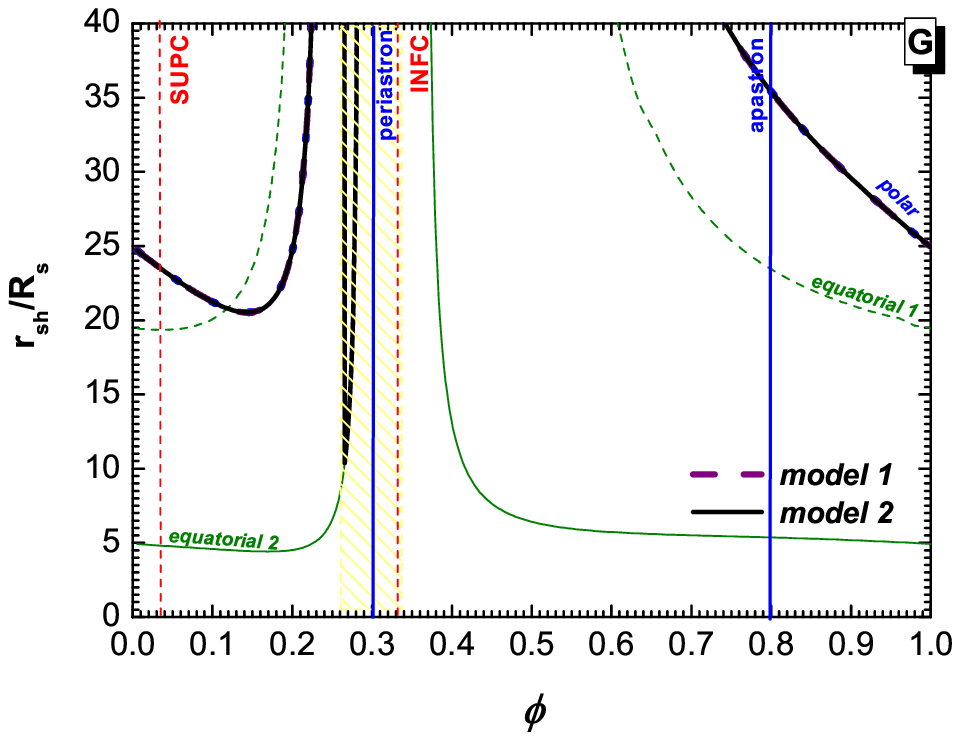}
   \includegraphics[width=0.48\textwidth,angle=0,clip]{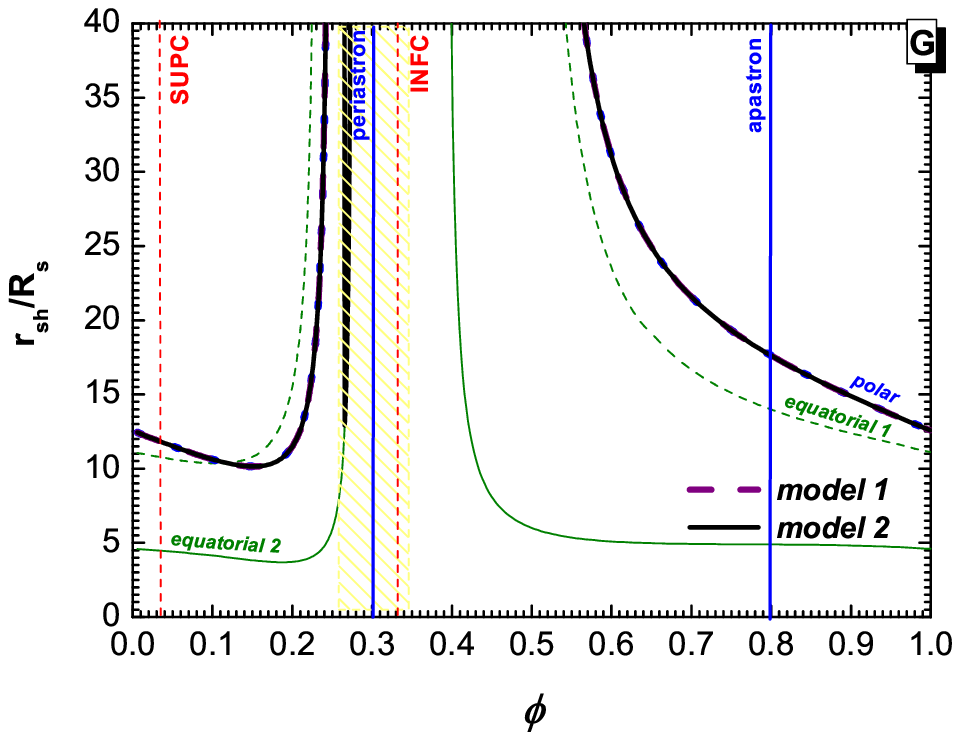}\\
   \includegraphics[width=0.48\textwidth,angle=0,clip]{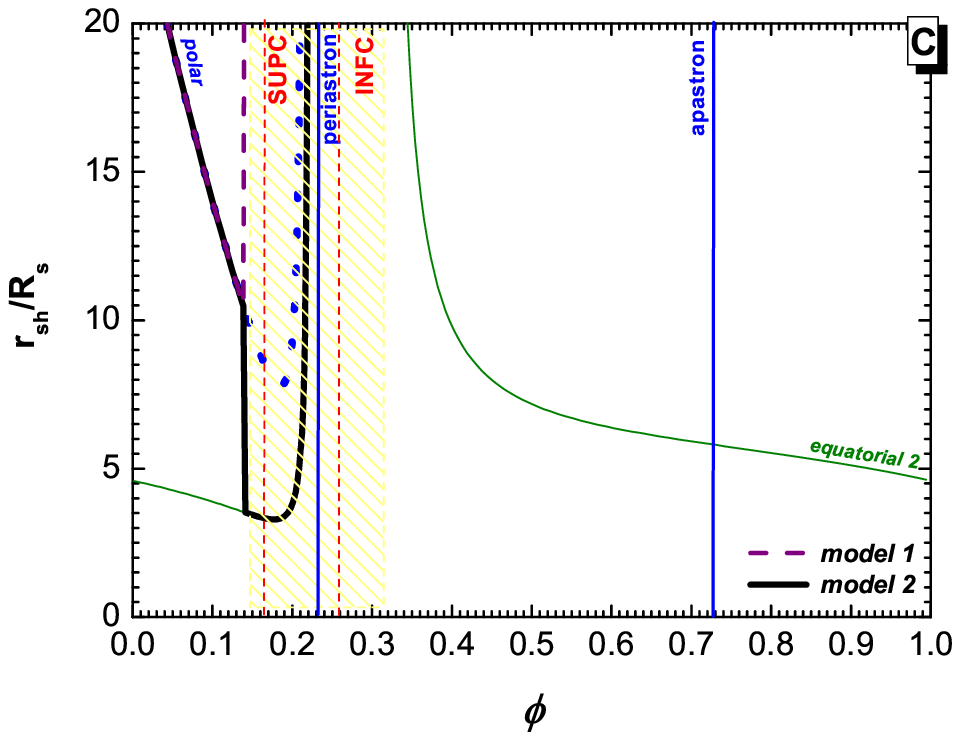}
   \includegraphics[width=0.48\textwidth,angle=0,clip]{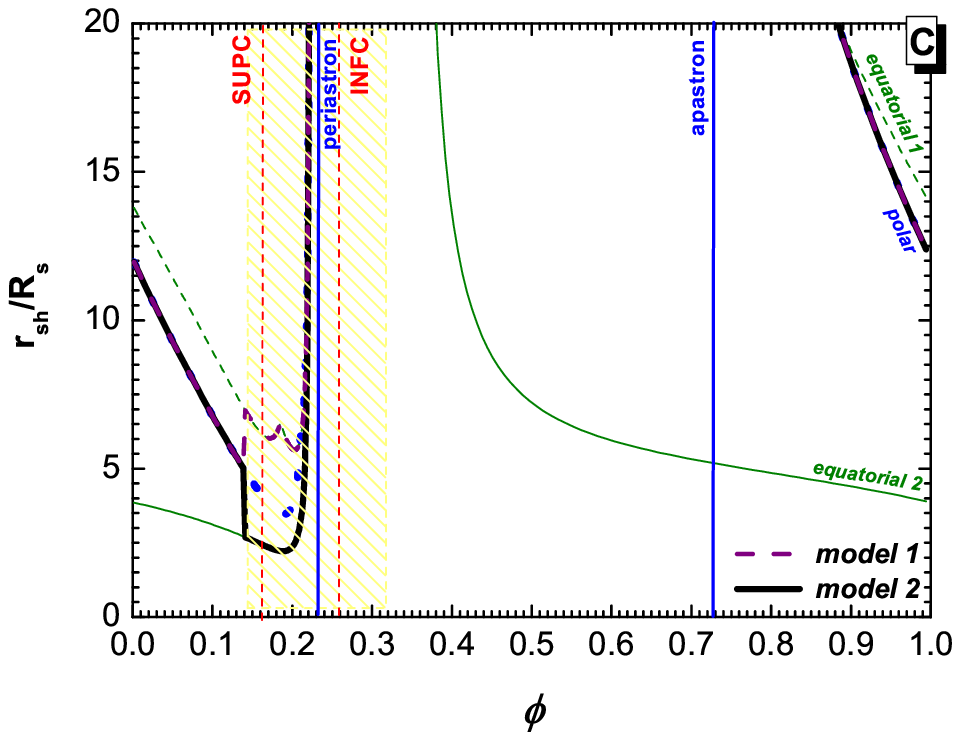}
\caption{\label{fig:shock_mixG} {The distance (in units of stellar radius $R_s$) from the pulsar  to the termination shock ($r_{sh}$) in the direction to the observer as a function of binary phase, for different Be wind models (where $r_d = 7 R_s$ and $\theta_d = 15^o$) and geometry based on Grundstrom et al. (2007, top, marked with G) and Casares et al. (2005, bottom, marked with C) and two inclination angles: $\textit{i} = 30^o$ (left), $\textit{i} = 60^o$ (right). The termination shock is marked with thick dashed line for model 1 and thick solid line for model 2 (Table \ref{tab:orb-param}). $r_{sh}$ was also calculated in the case of polar and equatorial winds only to facilitate visualising the components of model 1 and 2. Assuming the parameters of  the polar and equatorial winds as for isotropic wind the shock distance is marked by thin blue line for polar wind, dashed green (marked as ``equatorial 1'') line for 1 model with $f = 50$, and solid green (``equatorial 2'') line for $f = 500$. The results from polar wind parameters are covered most of the phases with the lines for mixed models. }
}
\end{figure*}

Note that for orbital parameters from Casares et al. 2005 (right panels), the separation at periastron is very small, and the values of $\eta$ are in fact not a good approximation for such geometry (the stand-off distance can be less than the distance from the massive star surface). In this case, the shock is assumed at a  minimal distance from the massive star, set at $1.3 R_s$ (see the dips around periastron already in the top-right panel of Fig. \ref{fig:etaGC}). This will also impact on the distance to the shock from the pulsar side, see Fig. \ref{fig:shock_mixG}.

For investigating the high energy photon production already in the PWZ, an important parameter is the distance from the pulsar to the front shock in the direction to the observer, $r_{sh}$, (see Sierpowska-Bartosik \& Torres 2007, 2008b). This magnitude, differently to the stand-off distance, depends on the orbital inclination. In Fig.  \ref{fig:shock_mixG} we show this distance along the orbit for both scenarios of the binary geometry, two inclinations (see the differences in angle to the observer in Fig. \ref{fig:alpha_obs}), and the two models considered for the stellar wind (different line styles in each plot). For comparison, the lines for different wind parameters (polar and equatorial) are plotted all along the orbit. The distance to the shock  for models 1 and 2 are  consistent with the values for isotropic polar wind (thin dotted blue lines in each figure) for phases when the pulsar is in its corresponding region of influence.

It is interesting to compare the phase periods when the PWZ is not terminated  in the direction to the observer  (i.e., an electron propagating in this direction will not find the shock) based on different models and geometries.
In Grundstrom et al. (2007) geometry (G) and within model 1, we have long period of a non-terminated PWZ  which changes with the system inclination: it is $\phi \sim 0.22-0.73$ for $i=30^o$, whereas it is $\phi \sim 0.24-0.56$ for  $i=60^o$, what comes from the interaction of the pulsar wind with the polar wind only (no termination shock in the equatorial wind; the efect is only in the direction to observer).
For the same scenario but within model 2, the equatorial wind is relatively stronger ($\eta < 1$) and the pulsar wind terminates in addition to previous phases in the disc region at short periods prior to periastron $\phi \sim 0.26-0.28$ ($i=30^o$) or $\phi \sim 0.26-0.27$ ($i=60^o$).
In the case of geometry based on Casares et al. (2006) we have shorter termination epochs. For $i=30^o$ in model 1  the wind is terminated only for $\phi \sim 0.04-0.14$ (what corresponds to the pulsar wind interaction with the polar wind). In model 2 this period becomes longer, $\phi \sim 0.04-0.22$, due to relatively stronger wind in eqatorial region. When the inclination is set to $i=60^o$ the magnitude of the angle to the observer is larger what results in shorter periods for the non-terminated pulsar wind. Then for both models, 1 and 2, we have the same phases when the wind is non-terminated, $\phi \sim 0.22-0.88$,  with differences in the value of $r_{sh}$. 
Note that in that last cases the radius $r_{sh}$ increases from few stellar radii to infinity just prior to the periastron.

\section{Opacities}

\begin{figure*} [th!]
  \centering
  \includegraphics[width=0.48\textwidth,angle=0,clip]{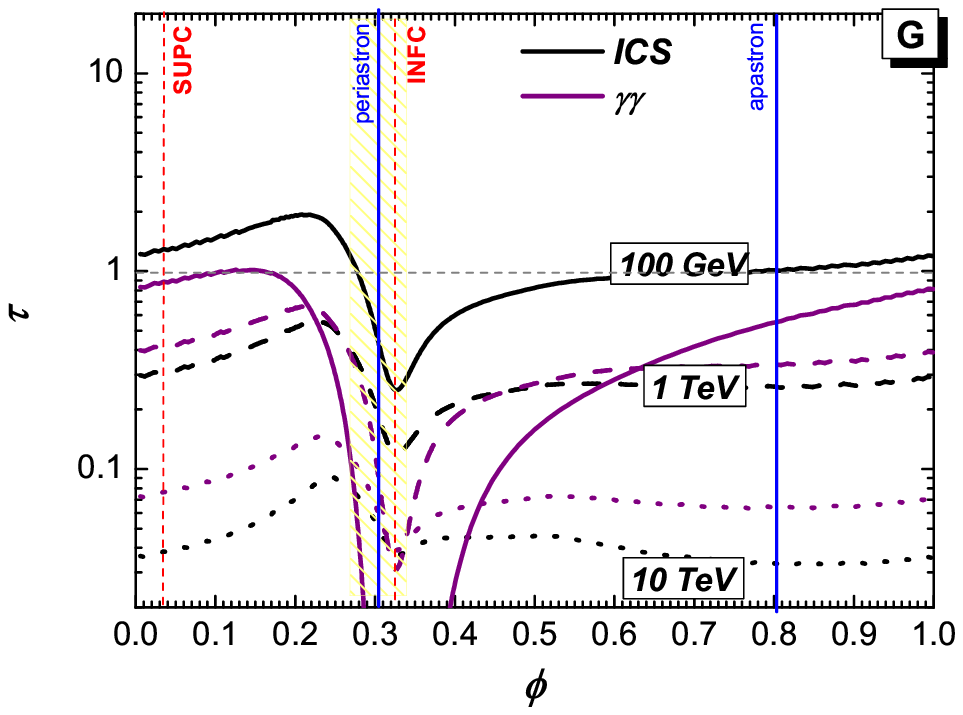}
   \includegraphics[width=0.48\textwidth,angle=0,clip]{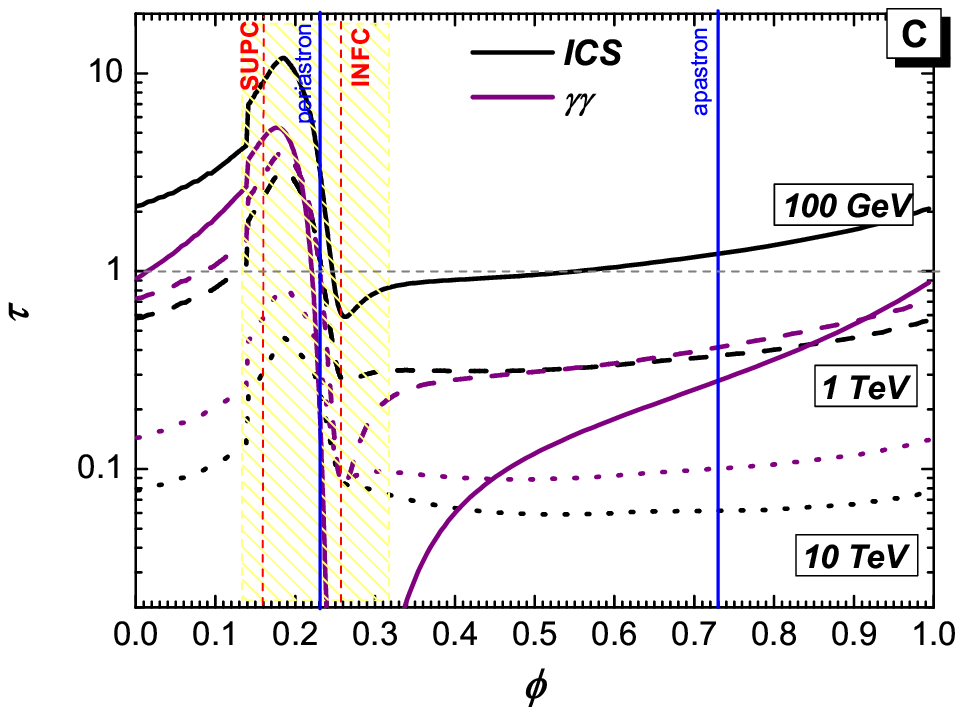}
\caption{\label{fig:tau_pwz_50} {The optical depths for Inverse Compton scattering (black lines) and $\gamma \gamma$ absorption (grey lines) calculated for particle injected at the binary separation distance and propagating up to the termination shock in the direction to the observer  for different primary energy. The optical depths were calculated for the model 1 with $f=50$ ($r_d = 7$,  $R_s$ and $\theta_d = 15^o$) and inclination $\textit{i} = 60^o$. The left panel correspond to the orbital solution by Grundstrom et al. (2007); the right one, by Casares et al. (2005).The electron/photon primary energy is set to $100$ GeV (solid lines), $1$ TeV (dashed lines) and $10$ TeV (dotted lines).}
}
\end{figure*}

The optical depths to inverse Compton interaction and $\gamma\gamma$ absorption were calculated for these models of the binary and its geometry to investigate high energy photon production. We used the full Klein-Nishina cross section { (see the Appendix of Sierpowska-Bartosik \& Torres 2008 for full details)}. We have computed the opacities for particles propagating to the observer from the position of the pulsar (assumed in the model to be the place of injection). We studied the influence of the different orbital elements (Grundstrom et al. and Casares et al.), the same inclinations of the orbit (60$^o$, in agreement with Hutchings \& Crampton (1981)  who actually favoured higher values of $i$), and assumptions on the stellar winds from model  1, see Fig. \ref{fig:tau_pwz_50}. { The results constitute a set of several plots that give account of the changes in opacities produced by uncertainties in our knowledge of fundamental parameters of the binary. }


Optical depths above unity are only found for low energies of pairs for inverse Compton, whereas for both pair production and inverse Compton processes  with higher initial energy, interactions are less probable. A high inclination in the orbital element solution provided by Casares et al. (2005) provide the highest opacities, as the angle to the observer reaches the maximum close to periastron (see SUPC position and also Fig. \ref{fig:alpha_obs})  where the radiation field density is the highest. This dependence of the angle to the observer along the orbit, together with the fact that in Casares et al. solution the system is more compact around periastron (larger radius of the star, larger eccentricity) than in Grundstrom et al.'s, results in higher optical depths for phases when the pulsar crosses the disc region. The termination of the pulsar wind do not influence the opacities much, as in the geometry of Grundstrom et al. and within model 1, the PSR wind is unterminated also in the disc region ($\phi \sim 0.24-0.56$), while in Casares et al.'s solution the wind is terminated up to phases close to periastron (termination for $\phi \sim 0.22-0.88$).
{ We find a similar behavior of the opacity curves between SUPC and INFC in both geometries, so the most important influences can be ordered as follows: geometry (angle to the observer + binary separation) - radiation field - and then the influence  from the shock termination (i.e., the jump at the disc we see in C comes from change of the $r_{sh}$ of the order of a few radii) }

Of course, the optical depths depend on energy so that, generally but not always, the top curves (larger opacities) correspond to the lower energy of injected particle (in these cases, $E$ = 100 GeV). The optical depths for pair production, in particular, 
present a peak in energy, which location depend on the angle of propagation. In this case, then, 
we may have lower optical depths for the lower energies explored- see the phases around periastron.
In summary, for the chosen particle energies: the optical depths for $E$ = 100 GeV on ICS are larger than those of pair production, for $E =$ 1 TeV they are both at a comparable level, and for $E =$ 10 TeV the optical depth for pair production are larger.

\begin{figure*}
   \centering
    \includegraphics[width=0.48\textwidth,angle=0,clip]{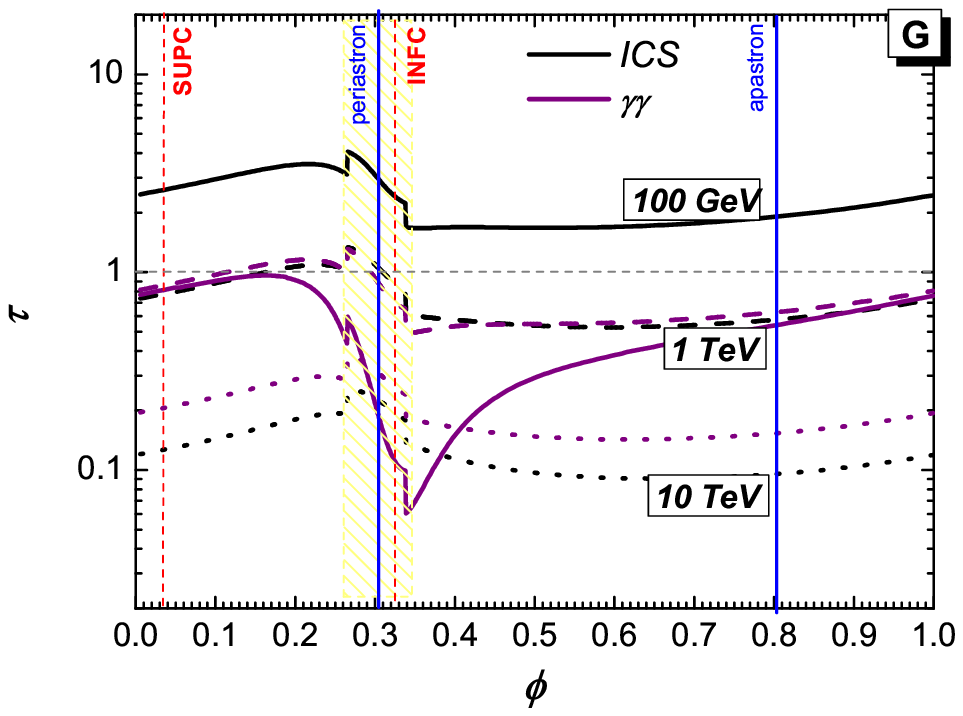}
    \includegraphics[width=0.48\textwidth,angle=0,clip]{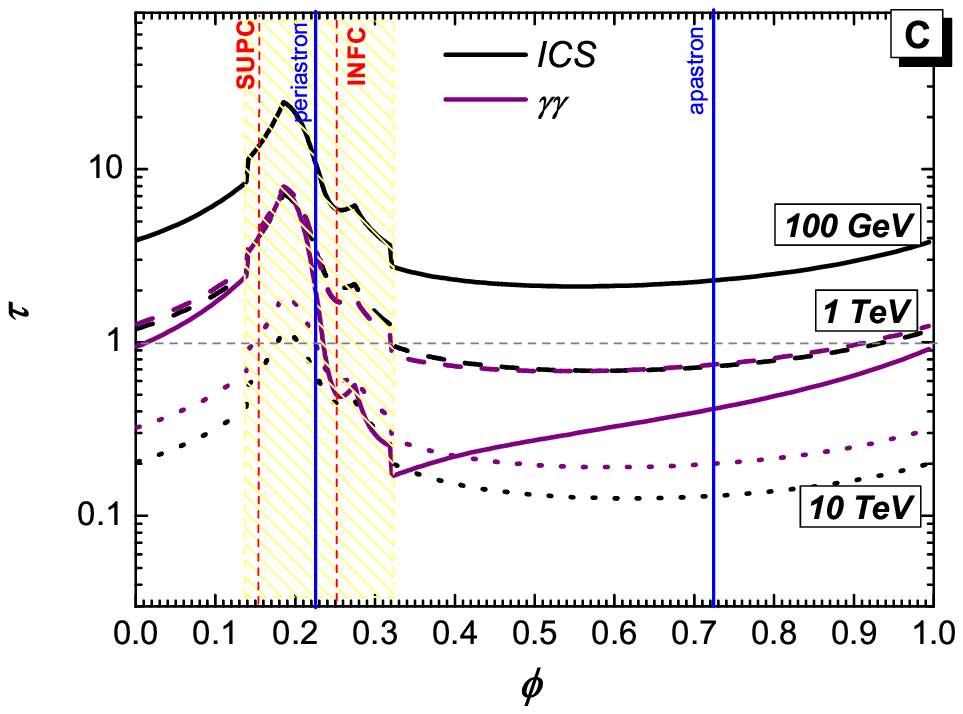}\\
   \includegraphics[width=0.48\textwidth,angle=0,clip]{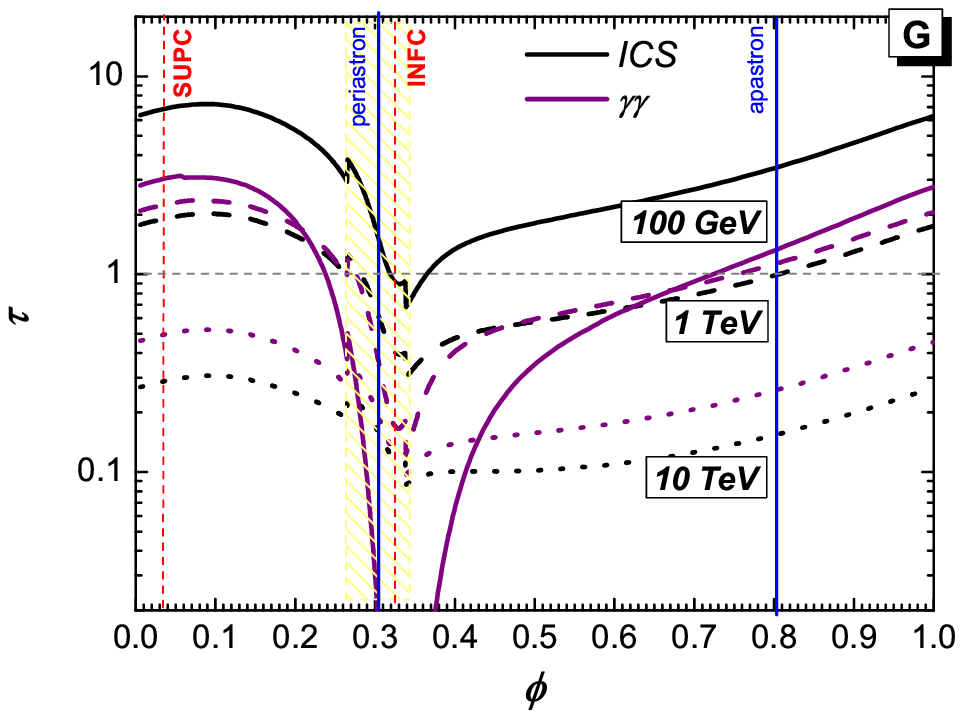}
    \includegraphics[width=0.48\textwidth,angle=0,clip]{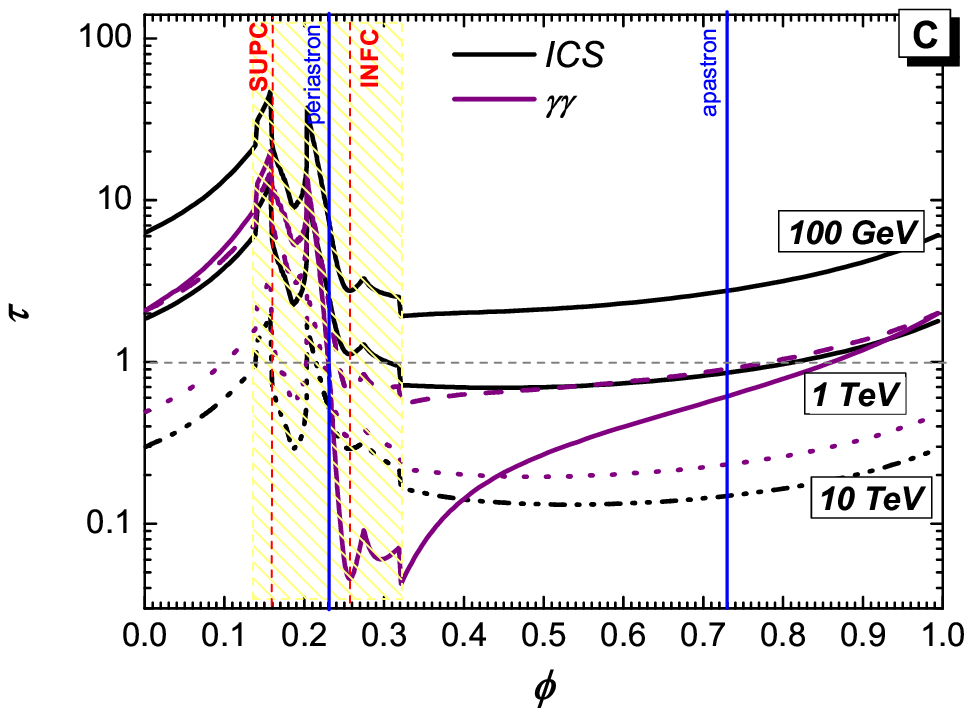}
 \caption{\label{fig:tau_sod_50} {The optical depths for Inverse Compton scattering (black lines) and $\gamma \gamma$ absorption (grey lines) calculated for particle injected at the stand-off distance and traveling in the direction to the observer for different primary energy. The optical depths were calculated for the model 1 with $f=50$ ($r_d = 7$,  $R_s$ and $\theta_d = 15^o$) and inclination $\textit{i} = 60^o$. The left panel correspond to using the orbital solution by Grundstrom et al. (2007); the right one, by Casares et al. (2005). }  }
 \end{figure*}

{ Notice that in a scenario where photons are produced by pairs accelerating along the shock, where magnetic field influence $e^\pm$ trajectories, photons are produced at different directions, not only at the specific observer's angle, and from different places of the shock.  This would require a full 3D treatment of the cascading process on which we will report elsewhere. But in order to describe 
the general dependencies of the emission to the densities and anisotropies local to the interaction regions, we present here the opacities in the direction to the observer at the specific phase and stand-off distance.

The optical depths at stand-off distance for model 1 are shown in Fig. \ref{fig:tau_sod_50}.  A more complicated scenario appears in the case of Casares et al. (2005) orbital solution (right panel of Fig. \ref{fig:tau_sod_50}), where the changes due to a minimal distance of the shock from the massive star for a limited-phase range around periastron, are seen. For this model, the optical depths for phases within the equatorial wind are higher as the stand-off distance is closer to the massive star. In the case of inclination 60$^o$, as we depict, there is a 
double-peak structure in the opacities, because for these limited phase-range, the directions to the observer (when starting from the stand-off distance) are eclipsed by the star surface; and in that cases the optical depth is lower (dip in the curve) since it is calculated only to the star surface. 
This double peak-structure is entirely geometrical, and is absent in the results for the opacities for  inclination 30$^o$, as is also in the case of the Grundstrom et al.'s (2007) orbital solutions. In this latter case, the influence of the disc is not significant mainly because the orbit occurs with a larger separation of the system. 

Only in case of Casares et al. (2005) parameters and high inclination values, the optical depths for both processes change significantly in the disc region with respect to the polar wind phases (the smallest and biggest angles of propagation happen for relatively small separations close to periastron). }
%


\begin{figure*}
   \centering
    \includegraphics[width=0.48\textwidth,angle=0,clip]{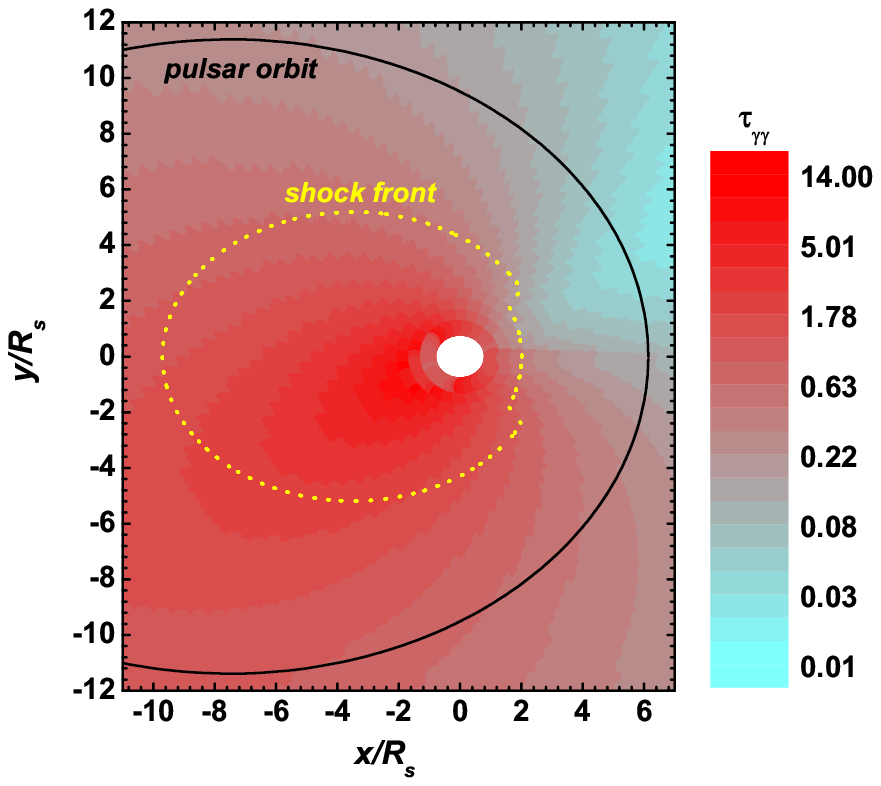}
    \includegraphics[width=0.48\textwidth,angle=0,clip]{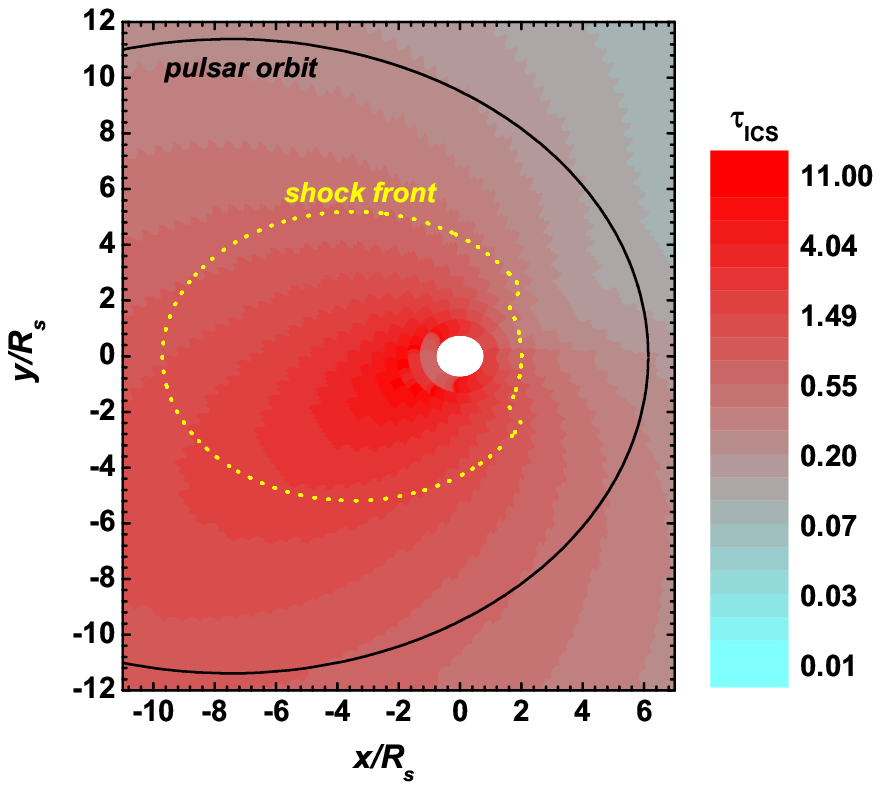}\\
 \caption{\label{fig:2Dmaps} {Optical depths for $\gamma \gamma$ absorption and Inverse Compton scattering calculated at all places in the orbital plane of the system, for 1TeV particles traveling in the direction to the observer. The Grundstrom et al. (2007) orbital solution is used. The orbit of the pulsar, the star (with its physical size), and the position of the shock are marked.  }}
 \end{figure*}

\begin{figure}
  \centering
  \includegraphics[width=0.48\textwidth,angle=0,clip]{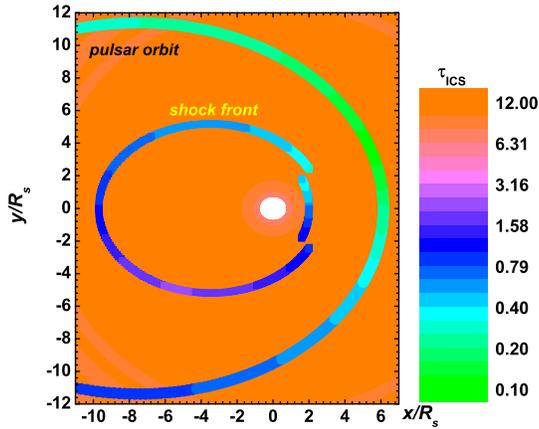}
 \caption{{\label{fig:2Dmaps2} Comparison of the for 1TeV particles maximal optical depths for Inverse Compton scattering calculated at all places in the orbital plane of the system (i.e., particles traveling in a direction tangent to the physical size of the star), with the optical depths found at the position of the shock and the pulsar orbit when particles are traveling in the direction to the observer. The Grundstrom et al. (2007) orbital solution is used.  The orbit of the pulsar, the star (with its physical size), and the position of the shock are marked.  }}
 \end{figure}

{ It is clear that the uncertainties in the system play a non-negligible role in the opacities. this in turn will translate into complexities of lightcurve and spectral evolution along the orbit that can be tested with future quality of data. For futher investigation we choose the geometry based on Grundstrom et al. (2007) and model 1 for the massive star wind, i.e., moderate values for the mass loss rates in the equatorial wind.

Fig. \ref{fig:2Dmaps} shows the optical depths for $\gamma \gamma$ absorption and Inverse Compton scattering calculated at all places in the orbital plane of the system, for 1TeV particles traveling in the direction to the observer. The orbit of the pulsar, the star (with its physical size), and the position of the shock are marked therein.  Fig. \ref{fig:2Dmaps2} shows something more complex yet: it presents a comparison of the for 1TeV particles maximal optical depths for Inverse Compton scattering calculated at all places in the orbital plane of the system (i.e., particles traveling in a direction tangent to the physical size of the star), with the optical depths found at the position of the shock and the pulsar orbit when particles are traveling in the direction to the observer. We see that the differences in optical depth produced by the direction of movement are clearly important; and thus the system geometry will decisively impact the lightcurve. }

%

\section{Application to a PWZ model of $\gamma$-production}

We apply these results to investigate the scenario for VHE photon production in the PWZ. As we have shown in previous papers (Sierpowska-Bartosik \& Torres, 2007,2008b) the geometry of the system is very important in this scenario because of, among other things, the linearity of particles' propagation (and further on cascading) from the injection place. 
As an example for \LSI, the PWZ model for the system was investigated based on the parameters of model 1 for geometry scenarios based on both orbital solutions, Casares et al. and Grundstrom et al.; in both cases, with an inclination fixed at $i=60^o$. 
Together with the velocities of the wind in the two regions as given in Table \ref{tab:orb-param}, this scenario is such that the interaction of the pulsar and massive star winds set the shock almost at half-separation distance (with $\eta$ close to 1), except for a very limited range of phases around the periastron where the shock is closer to the massive star, see Fig. \ref{fig:etaGC}, and consequently, where most of the difference between models are found. 

The $e^\pm$ pairs are assumed to come from within the PWZ, and we consider those in the direction to the observer, see Fig.\ref{fig:alpha_obs}. The electrons propagate linearly and interact with the soft photons from the massive star (anisotropic radiation) via inverse Compton. Photons following the same direction of the initial electron are thus created, which can initiate further pair production in the same photon field. These cascades are followed up to the termination shock, where the electrons are trapped by the local magnetic field but the photons pass by. Originated at the PWZ, photons can be absorbed in the massive star wind region (MSWR). All these processes are included in the simulations, see Sierpowska-Bartosik \& Torres (2007,2008, 2008b) for details. In this model, apart from the orbital solution and wind parameters, we need knowledge of the pair population interacting with the stellar photon field, as well as the normalisation factor (the amount of pulsar power transferred to primary pairs). We assume, based on both, observational results from high energy $\gamma$-rays (Albert et al. 2006, 2008a,b) and kinetic plasma physics studies in conditions found for the PWZ of pulsars (e.g., Jaroschek et al. 2008 and references therein) that this population is described by a power law. 
The fraction of the pulsar spin-down power ending in the $e^\pm$ interacting pairs can then be written as: $
\Gamma L_{sd} = \int N_{e^+e^-}(E) E dE.  
$
Assuming that the distance to the source is $d = 2.0$ kpc, the normalisation factor needed to compute the number of electrons travelling towards Earth is $ A = N_{e^+e^-}/ 4\pi d^2. $  In the case of a power-law in energy, $N_{e^+e^-}(E) \propto E^{-\alpha_i}$.
The photon spectra are simulated for $20$ phases chosen along the orbit of the binary (avoiding the phases close to the change between the polar and equatorial wind), which allows to compute the high energy lightcurve.

At TeV energies, the MAGIC observational lightcurve presents a broad maximum corresponding to phases from $0.5$ to $0.8$. Looking at Casares et al. and Grundstrom et al. geometry (Table \ref{tab:phase}) we can see that this corresponds to phases prior to apastron (and it is not connected with an INFC-SUPC distinction, as in the case of  LS 5039). Interestingly, this corresponds to phases where the PWZ in the direction to the observer is unterminated, and an increasing binary separation with the pulsar already outside the disc region. For the TeV maximum broad phases (between 0.5 and 0.8) we assume an initial spectrum of electrons given by a slope $\alpha = - 2.6$ and a normalization $\Gamma = 0.1$. The MAGIC data at other phases along the orbit are not particularly constraining, since in reality they represent only 95\% CL upper limits at high levels of flux (Albert et al. 2008b). 
To explore how this model can fit the MAGIC light curve (particularly if we assume very low values of fluxes at periastron) we change the parameters of the initial power law spectrum for electrons along the orbit up to $\alpha = -2$ and $\Gamma= 0.01$; implying, within the framework of this model, orbital variations in the relativistic population of particles, something already suggested in the case of LS 5039 (Sierpowska-Bartosik \& Torres 2008, Dubus et al. 2008). Needed variations in this case, are however stronger than those of LS 5039.

\begin{figure*}[t]
  \centering
   \includegraphics[width=0.51\textwidth,angle=0,clip]{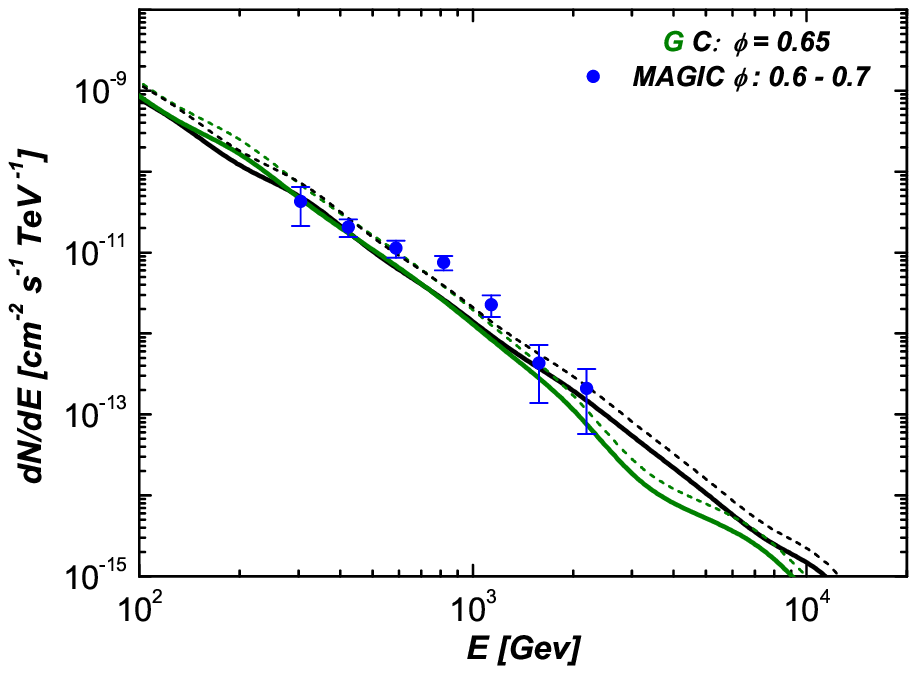}
   \includegraphics[width=0.48\textwidth,angle=0,clip]{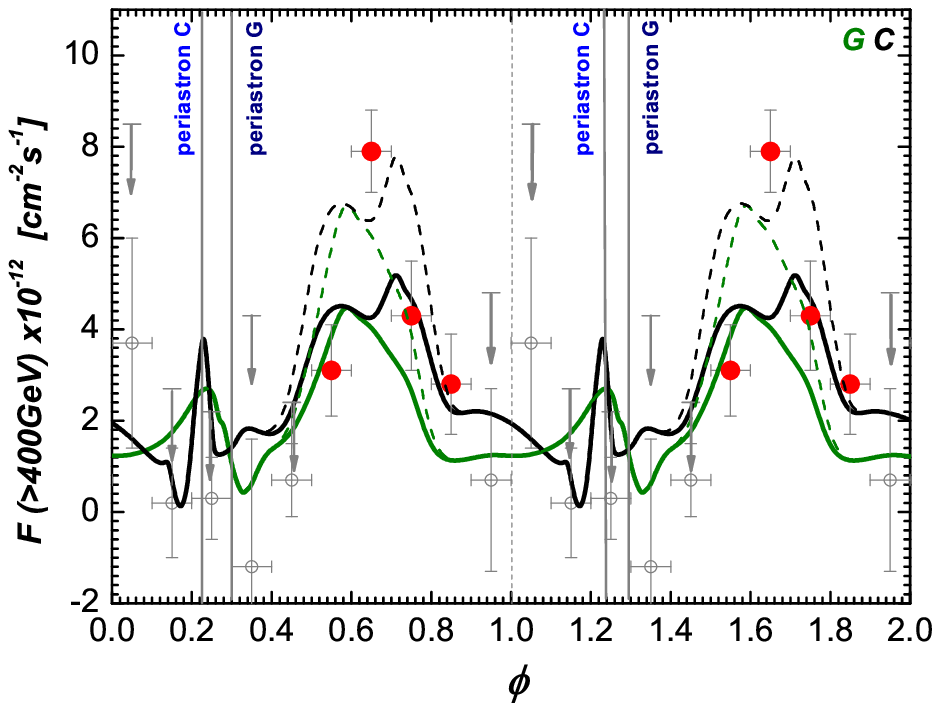}
\caption{\label{fig:lc_pwz} {The gamma ray spectrum  (where the TeV maximum occur, $\phi \sim 0.65$) and lightcurve  from interacting electrons in the PWZ, for system parameters based on the orbital solutions by Casares et al. (2005) - C and Grundstrom et al. (2007) - G. See the text for details. MAGIC data points are separated between those having less than 2$\sigma$ confidence (for which upper limits are also given) and higher confidence ones (shown as filled circles). Data is taken from from Albert et al. (2008b), error bars are statistical only. 
} }
\end{figure*}

Results for the two orbital models studied are qualitatively similar, but differ in the details, especially in what concerns to the lightcurve. These differences in the details are only coming from that produced by the uncertainty in the orbital solution elements, given that we have kept the same parameters in the rest of the variables (inclination, electron population, stellar wind, etc.). We show these results in Figure \ref{fig:lc_pwz}. In there, we also show with thin lines the results we would get for the phases of the central VHE maximum with a 5\% increase in normalization. Again, we emphasize that the results of the model at phases where there is no reliable detection are unconstrained, and are chosen to show a very low level of fluxes (larger fluxes are easy to achieve possible). Further constraints both by Fermi and MAGIC II in these orbital phases are essential to fix the properties of the model.


\begin{figure}
  \centering
   \includegraphics[width=0.48\textwidth,angle=0,clip]{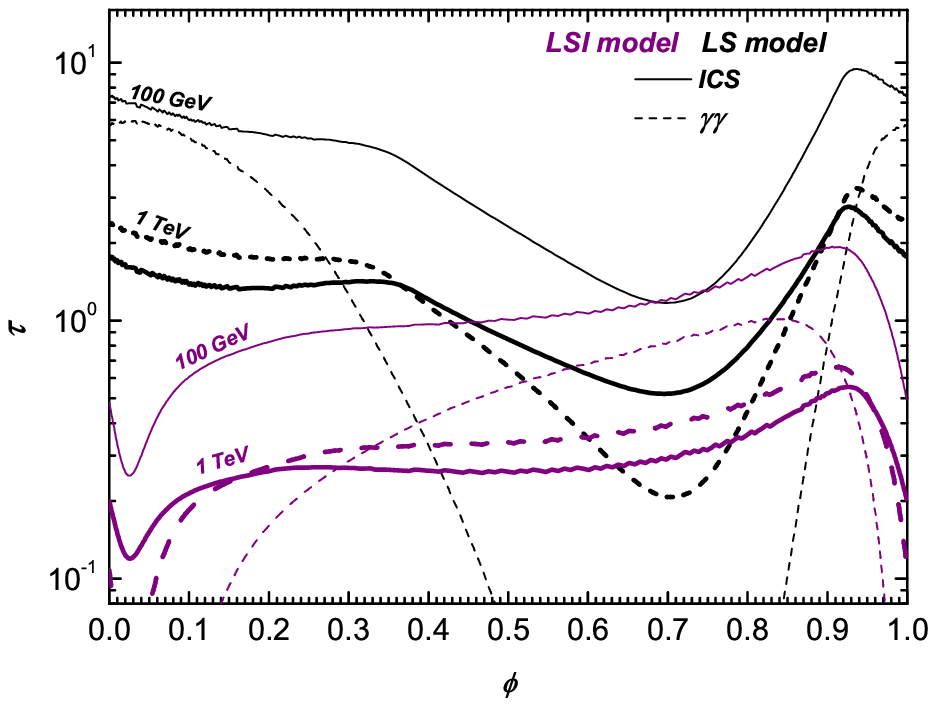}
\caption{\label{fig:tau_LSvsLSI} {Comparison of the optical depths in the PWZ of two different binaries: \LSI\ and LS 5039. Idem as in Fig. \ref{fig:tau_pwz_50} but only the model 1 was chosen for  \LSI\  with the orbital solution by Grundstrom et al. (2007) and the phase of periastron arbitrary set to $\phi =0$. In case of LS 5039 the parameters are given in Sierpowska-Bartosik \& Torres (2007).}
}
\end{figure}


\section{Concluding remarks}

This work shows that the range of uncertainties yet at hand in the orbital elements of the binary system \LSI, as well as in the possible basic assumptions on the stellar wind of the optical companion, 
play a non-negligible role in the computation of opacities to high energy processes leading to $\gamma$-rays. It is not only the face value of opacities what changes, but the geometrical influence on the propagation and escape of $\gamma$-ray photons is also different depending on the assumed parameters and orbital solutions. Detailed models of the binary needs to take these differences into account when making precise predictions for $\gamma$-ray fluxes. Hopefully, new observational studies will help to clarify at least some of the variables in the orbital solution.
As an example, we have shown that a conceptually simple PWZ model for the production of $\gamma$-ray emission (which details are discussed more at length in previous works) is qualitatively close to the data. If admitting orbital variations of the interacting electron populations along the orbit this model can also predict low level of fluxes for phases where MAGIC observations have up to now provided (yet non-constraining upper limits). Although this model is qualitatively in agreement with the MAGIC results,  it is, however, not complete. Particularly in the case of \LSI, we need to consider the acceleration of electrons at the shock front itself: {particularly for phases when opacities in the PWZ are not so high and many electrons will arrive there even if they are accelerated within the PWZ.}
Fig. \ref{fig:tau_LSvsLSI} presents a comparison of the corresponding optical depths in the \LSI\ system leading to the spectrum and lightcurve shown in Fig. \ref{fig:lc_pwz} with the opacities found for the LS 5039 model (Sierpowska-Bartosik \& Torres 2008b). To have a more direct comparison of the values we show the \LSI\ opacities after shifting the phase of periastron to $\phi = 0$.
 With no subsequent shock contribution, the fraction of spin-down power ending in relativistic electrons needed to fit the MAGIC TeV maximum (10\% $L_{sd}$) is larger than that found for LS 5039,  albeit still possible. Subsequent cascading in the MSWZ, if many electrons are reprocessed at the shock, can not be neglected. We expect to report on a more detailed model of gamma-production taking into account these effects elsewhere. { Future $\gamma$-ray observations (particularly at low energies, by Fermi) could help decide between these two models (PWZ and shock dominated) or explore their relative importance.}

\acknowledgements
We acknowledge  extended use of IEEC-CSIC parallel computers cluster.
We acknowledge N. Sidro and I. Ribas for discussions. 
This work was supported by grants AYA2006-00530,  AYA2008-01181-E/ESP, and CSIC-PIE 200750I029.

\appendix
\section{Appendix}

{  We present here the basic formulae for $\gamma\gamma$ opacity computations. Further details and plots (as well as the full treatment of inverse Compton interactions) can be found in the Appendix of Sierpowska-Bartosik \& Torres (2008).
The source of the thermal radiation field is the massive star of early type (O, Be, WR). The spectrum is described by Planck's law, which differential energy spectrum (the number of photons of given energy $\epsilon$ per unit energy $d\epsilon$, per unit solid angle $\Omega$, per unit volume $V$)  is given by:
\begin{equation}
\frac{dn(\epsilon,\Omega)}{d\epsilon d\Omega dV} = \frac {4
\pi}{(h c)^3} \frac{\epsilon^2}{e^{\epsilon /kT_{s}}-1}, 
\label{gp1}
\end{equation}
where $\epsilon$ is thermal photon energy, $h$ is the Planck constant, and $k$ is Boltzmann constant.


The optical depth to $\gamma$-photon absorption in the radiation field of the massive star up to infinity can then be calculated from the integral: 
\begin{equation}
\tau_{\gamma \gamma} (E_{\gamma},x_{i},\alpha) = 
\int_0^{\infty} \lambda_{\gamma \gamma} ^{-1}  (E_{\gamma},x_{i},\alpha,
x_{\gamma}) \, dx_{\gamma},
\label{gp14}
\end{equation}
where $\lambda_{\gamma \gamma}^{-1}$ is a photon interaction rate to  $e^\pm$ production in an anisotropic radiation field and $x_{\gamma}$  is its propagation length. When the propagation occurs toward the massive star surface, the integration is performed up to the stellar surface.
The photon interaction rate, $\lambda_{\gamma \gamma}^{-1}$, is related to a photon of energy  $E_{\gamma}$ injected at a distance $x_{i}$ from the massive star, at angle  $\alpha$,  and is given by the formula:
\begin{equation}
{\lambda_{\gamma \gamma}}^{-1} (E_{\gamma},x_{i},\alpha, x_{\gamma}) = \int
(1+\mu) d\mu
\int d\phi \int \frac{dn(\epsilon,\Omega)}{d\epsilon d\Omega dV} \sigma_{\gamma \gamma} (\beta) d\epsilon,
\label{gp2}
\end{equation}
where $x_{\gamma}$ is the distance to the interacting photon from the injection place along its propagating path.
The variable  $\mu$ in the first integration is the cosine of the photon-photon scattering angle $\mu = \cos \theta$. The angle $\phi$ is the azimuthal angle between the photon ($\gamma$-ray) propagation direction and the direction to the massive star, while  $\phi_s$ gives its limit value for the direction tangent to the massive star surface. 
The cross section to $e^\pm$ production is denoted as $\sigma_{\gamma \gamma} (\beta)$, where the parameter $\beta$ in the center of mass system is
$
\beta = \sqrt{\omega^2-m^2}/\omega, 
$
with 
$
\omega^2 = \frac{1}{2} E_{\gamma} \epsilon (1+\cos{\theta}) 
$
being the photon energy squared  (in this notation it is assumed also that $c=1$). 
From this we get 
$
\beta^2 = 1-{2m^2}/{E_{\gamma}\epsilon(1+\mu)}. 
$
The kinematic condition for the angle $\theta$ which defines the threshold for the $e^\pm$ creation is given by expression:
$
\mu \geq \mu_{lim} = \frac{2m^2}{E_{\gamma}\epsilon}-1.
$
To simplify the equations we rewrite the internal integration in Eq. (\ref{gp2}) making use of $I_1(\mu)$, such that,
\begin{equation}
I_1(\mu) = \int \frac{dn(\epsilon,\Omega)}{d\epsilon d\Omega dV} \sigma_{\gamma \gamma} (\beta) d\epsilon.
\label{gp3}
\end{equation}
With the replacement 
$
\beta^2 = 1 - a/\epsilon , 
$
where 
$
a=2 m^2/E_{\gamma}(1+\mu),
$ 
and defining the constant $S=8 \pi /(hc)^3$, the spectrum of thermal photons is now given by the formula:
\begin{equation}
\frac{dn(\epsilon,\Omega)}{d\epsilon d\Omega dV} = n(\beta) = S \frac{a^2}{(1-\beta^2)^2}
\frac{1}{e^{(a/(1-\beta^2)kT_{s})}-1}.
\label{gp4}
\end{equation}
Substituting in the internal integral $I_1(\mu)$, and introducing the integration variable to $\beta$, via 
$ 
d\epsilon = 2a \, \beta/(1-\beta^2)^2 \, d\beta, 
$
yields to the integral:
\begin{equation}
I_1(a) = 2S \int_0^1  \frac{\beta \, a^3}{(1-\beta^2)^4}
\frac{1}{e^{(a/(1-\beta^2)kT_{s})}-1} \sigma_{\gamma \gamma}(\beta) d\beta.
\label{gp5}
\end{equation}
The lower limit of integration is from the energy condition for the process $\gamma + \gamma \rightarrow e^+e^-$, i.e., it follows from the threshold condition $E_{\gamma} \epsilon (1+\mu) = 2 m^2$, where $\omega = m$. 
The upper integration limit comes from the relativistic limit $\omega \gg m$, where we get $\beta \approx 1$. 
To proceed forward, we introduce a dumb variable, $b$, by $a = kT_{s} b$, to get:
\begin{equation}
I_1(b) = 2S \int_0^1 (kT_{s})^3 b^3 \sigma_{\gamma \gamma}(\beta)
\frac{\beta}{(1-\beta^2)^4} \frac{1}{e^{(b/(1-\beta^2))}-1} \, d\beta.
\label{gp6}
\end{equation}


The cross section for $e^\pm$ pair production is:
\begin{equation}
\sigma_{\gamma \gamma}(\beta) = \frac{1}{2} r_0^2 \pi (1-\beta^2) \lbrack
(3-\beta^4) \ln \frac{1+\beta}{1-\beta} - 2\beta(2-\beta^2)\rbrack,
\label{gp7}
\end{equation}
where $r_0$ is the classical electron radius, and $\sigma_{T} = \frac{8}{3}
\pi r_0^2$ is the Thomson cross section. When putting this expression into the integral $I_1(b)$ (Eq. \ref{gp6}) we get finally,
\begin{eqnarray}
I_1(b) = C_1 \frac{3}{16} b^3 \int_0^1 \left[(3-\beta^4) \ln \frac{1+\beta}{1-\beta}
- 2\beta (2-\beta^2)\right]
\times \nonumber \\
 \frac{\beta}{(1-\beta^2)^3}
\frac{1}{e^{(b/(1-\beta^2))}-1} d\beta, 
\label{gp8}
\end{eqnarray}
with $C_1 = 16 \pi (kT_{s}/hc)^3 \sigma_T$.

We parametrize the internal integral and write it as a function of $b$, 
$ C(b) \equiv  I_1(\beta)/C_1.$
Then, the integral we are after can be written as $I_1(b) = C_1 C(b)$ and 
\begin{equation}
{\lambda_{\gamma \gamma}}^{-1} (E_{\gamma},x_{i},\alpha, x_{\gamma})  = C_1 \int
(1+\mu) d\mu \int d\phi \, C(b). 
\label{gp9}
\end{equation}
In the second internal integral, we have to fix the limits (having in mind that the angle $\phi$ depends on the angle $\theta$, then also on the variable $\mu$), 
\begin{equation}
I_2 = \int_{-\phi_{s}}^{\phi_{s}}C(b) d\phi = 2 \phi_{s} C(b)  =
C(b) \Phi(\mu). \label{gp10}
\end{equation}
{The angle $\phi_{s}$ determines the maximal azimuthal angle of $\gamma$-ray photon propagation with respect to the direction of the thermal photon, so that it gives the directions tangent to the star surface. This condition for $\phi_{s}$ can be determined from an spherical triangle} and the limits of integration with respect to the parameter $\mu$ are the range of angles for soft photons coming from the star (see Sierpowska-Bartosik \& Torres 2008 for details). Note that dependence on the distance to the massive star (given by the place of injection $x_{\gamma}$) is already in the calculation of the solid angle (parameters $\mu$ and $\phi$). When radius of the star is much smaller then the separation of the system $d$ (the place of photon injection) the integral over $\mu$ and $\phi$ gives the point source approximation (with the dependence $(R_s/d)^2$). \\
\\

}

\end{document}